\documentclass[12pt,preprint]{aastex}
\usepackage{graphicx}
\usepackage{subfigure}
\usepackage{multirow}

\begin{document}

\title{Probing the Rosette Nebula Stellar Bubble with Faraday Rotation}

\author{Allison H. Savage, Steven R. Spangler, and Patrick D. Fischer}
\affil{Department of Physics and Astronomy, University of Iowa,
   Iowa City, Iowa 52242}

\begin{abstract}

We report the results of Faraday rotation measurements of 23
 background radio sources whose lines of sight pass through or
 close to the Rosette Nebula.  We
 made linear polarization measurements with the Karl G. Jansky Very 
Large Array (VLA) at frequencies of 4.4 GHz, 4.9 GHz,
 and 7.6 GHz. We find the background Galactic contribution to the rotation measure in this part of the sky to be $+$147 rad m$^{-2}$.  Sources whose lines of
 sight pass through the nebula have an excess rotation measure of 
50-750 rad m$^{-2}$, which we attribute to the plasma shell of the Rosette 
Nebula. We consider two simple plasma shell models and how they reproduce the magnitude and sign of the rotation measure, and its dependence
 on distance from the center of the nebula. These two models represent 
different modes of interaction of the Rosette Nebula star cluster with the
 surrounding interstellar medium. Both can reproduce the magnitude and spatial extent of the rotation measure enhancement, given plausible free parameters. We contend that the model based on a stellar bubble more closely reproduces the observed dependence of rotation measure on distance from the center of the nebula.

\end{abstract}


\section{Introduction}

Luminous young stars interact with and alter the Interstellar Medium (ISM) from which they form. They interact by photoionizing gas in their vicinity, leading to a propagating ionization front \citep{spi1968}, and by the powerful stellar winds formed by hot, luminous stars. Over the course of a stellar lifetime, stellar winds modify the ISM by inflating a bubble of hot gas surrounding a star cluster. The \citet{wea1977} solution for the bubble due to a single star consists of an inner termination shock, a surrounding bubble of hot, low density stellar gas, a contact discontinuity, interstellar medium gas that is photoionized, and finally, an outer shock through which the interstellar gas has passed. A diagram illustrating this structure is given in Figure 1 of \citet{fre2003}. Within this picture, the visible HII region corresponds to the annular shell of shocked, photoionized gas. Whether a bubble structure exists, or instead a less dynamic structure corresponding to an ionization front, depends on the mechanical luminosity of the wind or winds in the star cluster. The long term goal of our research program is to better understand how stars in OB associations modify the ISM. In this paper, we present results on Faraday rotation measurements (a diagnostic of plasma properties) on lines of sight through the ionized ``bubble'' produced by one OB association, and interpret the measurements in the context of models of young clusters.

HII regions are plasmas, and principles of plasma physics determine how these structures evolve and impact the surrounding interstellar medium. One of the most important properties of an astrophysical plasma is the magnetic field. The magnetic field in an HII region or stellar bubble can strongly impact the evolution of the HII region or bubble. At the same time, modification of the magnetic field in the vicinity of an HII region could have consequences for subsequent star formation, properties of interstellar turbulence, and heat flow, among other processes. Measurement of magnetic fields in the interstellar medium is notoriously difficult. One of the best available techniques,  and the one utilized in this paper, is Faraday rotation of linearly polarized radio waves from extragalactic radio sources (described in Section 1.1 below; see \citet{min1996, hav2004, hav2006, bro2003, bro2007, val1993, val2004}, among others, for prior uses of this technique). An attractive aspect of Faraday rotation is that it can also be measured for lines of sight that pass through the solar corona, and thus provide information on the coronal magnetic field \citep{man2000, ing2007}. The fact that the same diagnostic technique can be used in these two media may facilitate comparison between plasma processes in the corona and solar wind, and those in the interstellar medium.

The specific object for study in this paper is the Rosette Nebula, which is a prominent HII region featuring an obvious shell structure and a central cavity (see Figure \ref{figrossource}). It is located on the edge of a molecular cloud in the constellation Monoceros. We adopt as its center that of the NGC 2244 star cluster (which is responsible for the Rosette), which is given by \citet{ber2002} as RA(J2000)= 06$^h$ 31$^m$ 55$^s$, Dec(J2000)=04$^o$ 56' 34" ($l$=206.5, $b$=-2.1). The distance to the Rosette is 1600 parsecs and its age is estimated to be 3 $\pm$ 1 Myr old \citep{Rom2008}.  \citet{men1962} concluded that the Rosette Nebula is an ionization-bounded Str\"{o}mgren sphere on the basis of radio continuum observations and that its structure is that of an annular shell. This structure is consistent with that of a wind-blown bubble, as mentioned above.

Within the central cavity of the Rosette is the OB stellar association NGC 2244. Photometry and spectroscopy studies put the age of NGC 2244 at less than 4 Myr \citep{per1989}. Evolutionary models place the main-sequence turn off age at 1.9 Myr \citep{Rom2008}. Despite this age discrepancy, both theoretical models and observations indicate that NGC 2244 is still forming stars \citep{Rom2008}. There are 21 confirmed pre-main sequence stars, and 113 confirmed stars belonging to NGC 2244, of which at least 7 are O type stars and 24 are B type stars \citep{Rom2008, park2002, ogu1981, wan2008}. The two brightest stars are HD 46223, an O4V star, and HD 46150, an O5V star \citep{Rom2008, wan2008}.

\subsection{Faraday Rotation as a Diagnostic Technique for Stellar Bubbles}

Faraday rotation is an excellent diagnostic tool for estimating properties of astrophysical plasmas such as the density of the general interstellar medium and the large scale structure of the Galactic magnetic field. Faraday rotation is the rotation in the plane of polarization of a radio wave as it propagates through a plasma that has a magnetic field. The polarization position angle $\chi$ of a source, or part of a source, whose radiation has propagated through the ISM is given by

\begin{equation}
\chi=\chi_{0}+\left[\left(\frac{e^{3}}{2\pi m_{e}^{2}c^{4}}\right)\int_0^{L} {n_e\vec B\cdot \vec{ds}}\right]\lambda^{2}
\label{RM1}
\end{equation}
where $\chi$ is the polarization position angle, $\chi_{0}$ is the intrinsic polarization position angle (i.e. that which would be measured in the absence of a medium), $\emph{e}$ is the fundamental electric charge, $\emph{$m_{e}$}$ is the mass of the electron, $\emph{c}$ is the speed of light, $\emph{n$_e$}$ is the electron density, $\emph{$\vec {B}$}$ is the magnetic field, $\emph{$d\vec{s}$}$ is the incremental pathlength interval along the line of sight, and $\emph{$\lambda$}$ is the wavelength. The integral in Equation (1) is taken from the source at $s=0$ to the observer at $s=L$. The variable $L$ represents the effective thickness of the plasma. With this convention, a positive value for the integral corresponds to the average magnetic field pointing from the source to observer, while a negative value represents a mean magnetic field pointing from the observer to the source. The quantity in square brackets is defined as the rotation measure (RM). The fundamental definition of Faraday rotation given in Equation (\ref{RM1}) is in cgs units. Values of RM are conventionally given in SI units. This conversion can be accomplished by multiplying the cgs value of the RM by a factor of $10^4$ to obtain the SI value. Alternatively, an expression which gives an SI value for the RM given mixed but convenient interstellar units  is \citep{min1996}
\begin{equation}
RM=0.81\int^L_0 n_{e} (cm^{-3}) \vec{B}(\mu G)\cdot \vec{ds} \mbox{ (pc)  rad m$^{-2}$ }
\label{RMSI}
\end{equation}
 Equation (\ref{RM1}) shows that if measurements of $\chi(\lambda)$ are available at two or more wavelengths (preferably three or more), the RM can be measured as the slope of a line through the data on a plot of $\chi$ vs. $\lambda^{2}$,
\begin{equation}
RM=\frac{\Delta\chi}{\Delta(\lambda^{2})}
\label{RM}
\end{equation}
The wavelengths of observation must be spaced closely enough that there is no possibility of a ``wrap'' of $\pi$ radians between two adjacent frequencies of observation.  This is referred to as the ``n-$\pi$ ambiguity''.  A discussion of the constraints on spacing between observing frequencies, as well as an illustration of the difficulties if they are spaced too far apart, is given in \cite{laz1990} (see Figures 3 and 4 of that paper).  Further details of how we extract RM values from our data are given in Section 3.2.

Among the many studies to have used Faraday rotation in the investigation of interstellar magnetic fields are \citep{rand1989,min1996,bro2003,bro2007,har2011,van2011}. To extract information on the magnetic field, it is necessary to have information on the electron density, since the integrand in Equation (\ref{RM1}) is the product of $\emph{n$_{e}$}$ and B$_{||}$, the parallel component of the interstellar magnetic field. The data sources we use for estimates of $n_{e}$ are described in detail in Section 4.1 below.

\subsection{The Rosette Nebula as a Candidate for Faraday Rotation Measurements}

The Rosette Nebula is an excellent object for studies of stellar bubbles via the technique of Faraday rotation. Besides being a prominent HII region with a shell and cavity, the Rosette has other properties which make it an excellent choice for studies of the impact of a young stellar association on the surrounding ISM. The Rosette is in the rough direction of the Galactic anticenter ($\emph{l}$=206.5$^o$). This gives it a number of advantages relative to HII regions and young star clusters in the inner two quadrants of the Galactic plane. Since star formation regions are relatively rare beyond the solar circle, there is no confusion in the Rosette field with other star formation regions at different distances along the line of sight. By contrast, studies in the Cygnus Region (e.g. \citet{whi2009}) are complicated by numerous star formation regions at various distances. Extinction also is less heavy for most anticenter lines of sight. The star cluster responsible for the Rosette Nebula (NGC 2244) is clearly seen, and the spectral types of the stars have been determined.

Another advantage of the Rosette Nebula is its structural simplicity. It resembles the theoretical ideal of a photoionized interstellar bubble as described by the theory of \citet{wea1977}. Furthermore, the parameters of the bubble structure have been determined by the radio continuum observations of \citet{men1962}, and later confirmed by \citet{cel1983,cel1985}. \citet{cel1985} determined that the Rosette Nebula is a spherical shell of ionized matter around NGC 2244 on the basis of radio continuum observations at 1.4 GHz and 4.7 GHz with the 100 m telescope at Effelsberg. Celnik also reported values for the inner and outer radius of the shell of gas and the density within the HII region \citep{cel1985}\footnote{\footnotesize{See Celnik 1985, Section 5.1 for the details of those measurements.}}. We adopt Celnik's parameters for the shell density and the structure in our analysis in Section 4.

\subsection{Previous results of Faraday Rotation Diagnostics of HII Regions}

\citet{whi2009} presented a study of the Galactic plane region near the Cygnus OB1 association. The main purpose of \citet{whi2009} was to confirm the existence of a ``Faraday Rotation Anomaly'' in this part of the sky, i.e., a large change in RM over a small distance on the sky. \citet{whi2009} argued that this anomaly was due to the plasma bubble associated with the Cygnus OB1 association. \citet{whi2009} also developed a simple shell model that reproduced the observed magnitude and the change in RM in Cygnus. In Section 4, we will use this shell model to interpret our data on the Rosette Nebula.

\citet{har2011} used Faraday rotation and H$\alpha$ measurements with the WHAM  spectrograph \citep{haf2003} to measure the electron density and line of sight magnetic fields in several HII regions. Faraday rotation was measured for extragalactic radio sources viewed through the HII regions. They probed 93 lines of sight in 5 HII regions, and found that each HII region displays a coherent magnetic field, with a range of 2 to 6 $\mu$G for the parallel component \citep{har2011}. \citet{har2011} briefly compared their RM values with the model presented by \citet{whi2009} and concluded that there is no evidence for a shell with an amplified magnetic field in any of the HII regions. \citet{whi2009} and \citet{har2011} thus come to different conclusions about the nature of the plasma shell that comprises an HII region. It should be noted that \citet{whi2009} claimed that the Faraday rotation anomaly was consistent with a wind-blown bubble, but did not claim that it was inconsistent with a shell without magnetic field amplification. Additional observations of the sort presented in \cite{whi2009} will help resolve this issue.  Measurements of $RM$ on a large number of lines of sight through an HII region (in the case of the present paper, the Rosette Nebula) will diagnose the plasma structure of the HII region, and determine if the HII region produces significant modification of the interstellar magnetic field.  In time, we plan to carry out such observations on a set of HII regions associated with star clusters of different age, stellar luminosity, and wind power.

\section{Observations}

\begin{figure}[ht!]
\begin{center}
\includegraphics[scale=0.12]{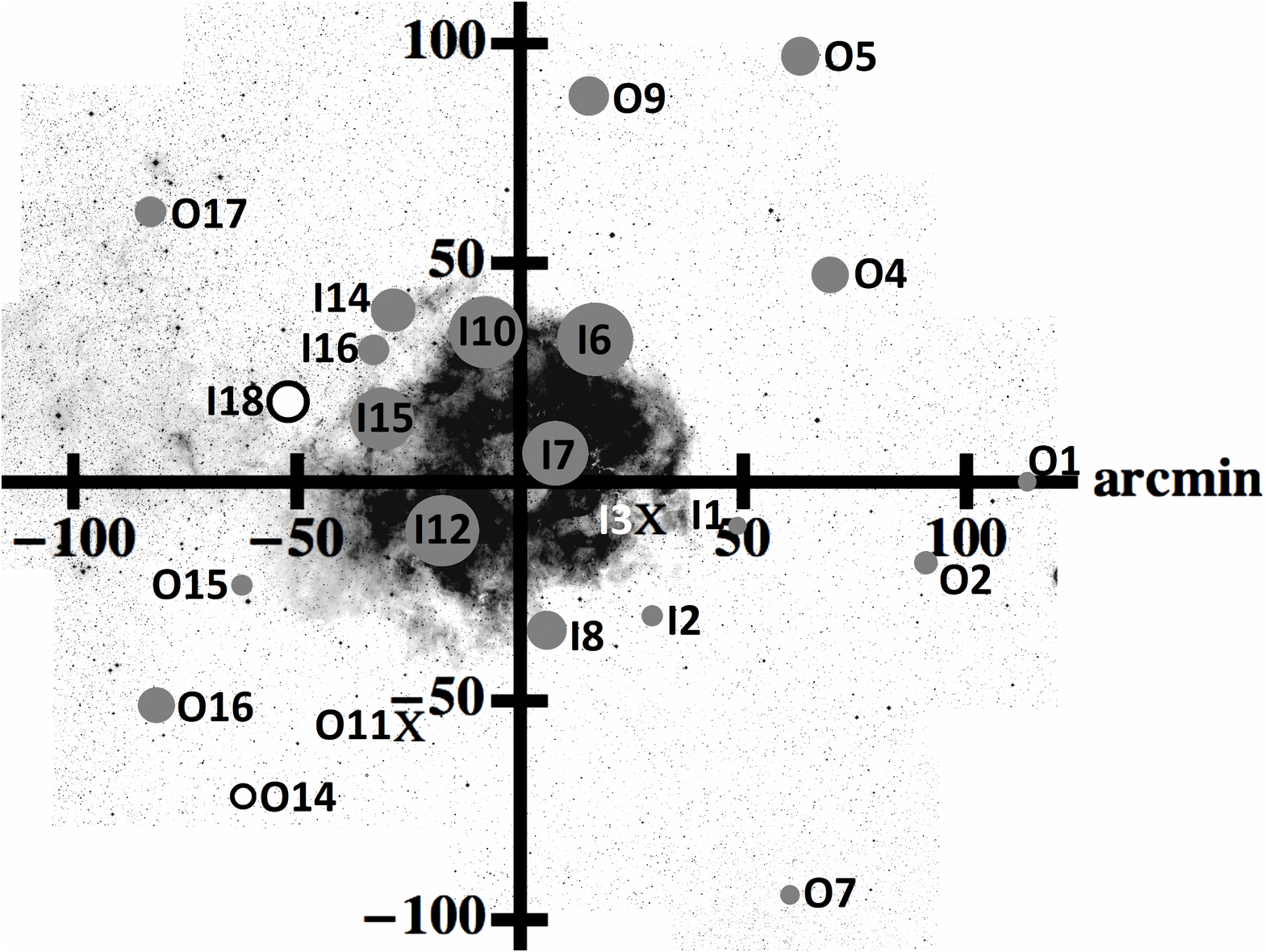}
\caption{A mosaic of the Rosette Nebula compiled from the Palomar Sky Survey II. The interior sources whose lines of sight pass through, or close to, the visible nebula are labeled with the prefix of ``I''. The exterior sources  whose lines of sight are well outside the visible nebula are labeled with the prefix of ``O''. Sources with negative RMs are labeled with open circles and those with positive RMs have solid circles. Depolarized sources are marked with an ``X''. The source symbols are scaled with the magnitude of the log $\mid$RM$\mid$.  \footnotesize{[The Second Palomar Observatory Sky Survey (POSS-II) was made by the California Institute of Technology with funds from the National Science Foundation, the National Geographic Society, the Sloan Foundation, the Samuel Oschin Foundation, and the Eastman Kodak Corporation. The STScI Digitized Sky Survey can be found at [http$://$stdatu.stsci.edu$/$cgi$-$bin$/$dss$_{}$form.]}}
\label{figrossource}
\end{center}
\end{figure}

All observations were made with the Karl G. Jansky Very Large Array (VLA) radio telescope of the National Radio Astronomy Observatory during the first several months of commissioning of the upgraded VLA.\footnote{\footnotesize{The Very Large Array is an instrument of the National Radio Astronomy
Observatory. The NRAO is a facility of the National Science Foundation,
operated under cooperative agreement with Associated Universities, Inc.}} Details of the observations and resultant data are given in Table \ref{tbl2}. The VLA was in D array for all of the observations. We observed 23 extragalactic radio sources whose lines of sight pass through or close to the Rosette Nebula. The sources were chosen from the NRAO VLA Sky Survey (NVSS), which covers the entire sky north of declination -40$^{\circ}$  at 1.4 GHz \citep{con1998}. We also observed four calibrators, 3C286, J0632+1022, J0643+0857, and 3C138. The calibrator 3C286 is commonly used for absolute calibration of the visibility amplitudes because it has a  well known flux density. It is also used to calibrate the origin of the polarization position angle. The source 3C138 was used for independent observations that could also set the flux density scale and determine the origin of the polarization position angle. Specifically, we used our observations of 3C138 to independently confirm the value of the R-L phase difference (used to calibrate the polarization position angle) obtained from 3C286. The source J0632+1022 was the primary calibrator for the project, functioning as the gain calibrator, i.e., determining the complex gain of each antenna as a function of time. This source (J0632+1022) was also used to measure the instrumental polarization, described by the ``D factors'', D$_R$ and D$_L$ \citep{big1982, sak1994}. We also observed a second source, J0643+0857, to obtain a completely independent set of D factors which confirmed our instrumental polarization calibration.

In addition to the calibrators, we observed 23 program sources. We had 12 sources whose lines of sight passed through the Rosette Nebula. The remaining 11 sources have lines of sight that pass near the Rosette Nebula but outside the obvious H$\alpha$-emitting shell. We observed these latter sources so we could establish a background RM value due to the Galactic plane. Figure \ref{figrossource} shows an image of the Rosette Nebula with the positions of our sources superposed. 

\begin{table}[!ht]
\begin{center}
\caption{Log of Observations \label{tbl2}}
\begin{tabular}{ll}
\tableline\tableline
Dates of Observation & March 20, 2010; July 4, 2010; August 22, 2010\\
Duration of Observing Sessions (h) & 5.95; 5.89; 5.94\\
Frequencies of Observations\tablenotemark{a}(MHz) &4136; 4436; 4936; 7636\\
VLA array & D\\
Restoring Beam (diameter) & 12\farcs8; 19\farcs6\tablenotemark{b}\\
Number of Scans per Source Per Session & 5\\
RMS Noise Level in Q and U Maps (mJy/Beam) & 0.042; 0.048; 0.037\tablenotemark{c}\\
\tableline
\end{tabular}
\tablenotetext{a}{\footnotesize{The observations had 128MHz wide spectral windows centered on the frequencies listed.}}
\tablenotetext{b}{\footnotesize{The March and July sessions were restored using a beam size of 12\farcs8, and the August session was restored with a 19\farcs6 beam size.}}
\tablenotetext{c}{\footnotesize{Average RMS noise levels for 4.4 GHz, 4.9 GHz, and 7.6 GHz, respectively.}}
\end{center}
\end{table}

\clearpage
\begin{table}
\begin{center}
\tabletypesize{\scriptsize}
\caption{Sources Observed. \label{sources}}
\begin{tabular}{cccccccc}
\tableline\tableline
Source & $\alpha$(J2000) & $\delta$(J2000) & $\emph{l}$  & $\emph{b}$ & $\xi$\tablenotemark{a} & S(4.9GHz)\tablenotemark{b} & Number of \\
Name & h m s & $^o$ ' '' &  ($^o$) & ($^o$) & (arcmin) & [Jy] & f observed \\
\tableline
I1 & 06 28 39.50 & 04 47 08.0 & 206.1 & -2.9 & 49.6 & 0.017 & 2\\
I2 & 06 29 56.26 & 04 26 33.0 & 206.5 & -2.7 & 42.2 & 0.260 & 3 \\
I3 & 06 29 57.30 & 04 47 45.5 & 206.2 & -2.6 & 30.6 & 0.038 & 3\\
I6 & 06 30 50.04 & 05 29 26.6 & 205.7 & -2.1 & 36.6 & 0.013 & 2\\
I7 & 06 31 24.28 & 05 02 50.8 & 206.2 & -2.1 & 9.9 & 0.043& 2\\
I8 & 06 31 34.31 & 04 22 34.4 & 206.8 & -2.4 & 34.4 & 0.025 & 3\\
I10 & 06 32 31.12 & 05 30 32.7 & 205.9 & -1.7 & 35.2 & 0.024 & 2\\
I12 & 06 33 03.14 & 04 44 56.0 & 206.6 & -1.9 & 20.6 & 0.047 & 3\\
I14 & 06 33 46.34 & 05 36 54.0 & 205.9 & -1.4 & 48.9 & 0.070 & 3\\
I15 & 06 34 00.01 & 05 10 42.8 & 206.3 & -1.5 & 34.2 & 0.021 & 3\\
I16 & 06 34 11.48 & 05 25 32.0 & 206.1 & -1.3 & 44.7 & 0.020 & 2\\
I18 & 06 35 25.96 & 05 14 15.3 & 206.4 & -1.2 & 55.4 & 0.028 & 2\\
O1 & 06 24 18.84 & 04 57 01.9 & 205.4 & -3.7 & 113.6 & 0.150 & 2\\
O2 & 06 25 51.89 & 04 35 40.2 & 205.9 & -3.6 & 92.8 & 0.340 & 3\\
O4 & 06 27 21.09 & 05 45 37.8 & 205.1 & -2.7 & 84.0 & 0.090 & 3\\
O5 & 06 27 36.73 & 06 32 52.1 & 204.4 & -2.3 & 115.7 & 0.066 & 3\\
O7 & 06 27 38.32 & 03 24 59.6 & 207.2 & -3.7 & 111.7 & 0.220 & 2\\
O9 & 06 30 52.53 & 06 24 50.5 & 204.9 & -1.6 & 89.6 & 0.050 & 3\\
O11 & 06 33 32.77 & 04 00 06.0 & 207.3 & -2.1 & 61.5 & 0.110 & 2\\
O14 & 06 35 51.95 & 03 42 18.0 & 207.9 & -1.8 & 94.9 & 0.029 & 2\\
O15 & 06 36 05.69 & 04 32 40.5 & 207.1 & -1.3 & 66.9 & 0.410 & 3\\
O16 & 06 37 23.05 & 04 05 44.1 & 207.7 & -1.3 & 96.3 & 0.029 & 3\\
O17 & 06 37 36.18 & 05 55 32.5 & 206.1 & -0.4 & 103.4 & 0.038 & 2\\
\tableline
\end{tabular}
\tablenotetext{a}{Angular distance between the line of sight and a line of sight through the nebula. See Section 4.1.}
\tablenotetext{b}{Total flux at 4.9GHz.}
\end{center}
\end{table}

We observed 128 MHz wide spectral windows centered on three frequencies: 4.436 GHz, 4.936 GHz, and 7.636 GHz. We had three sessions on the VLA (``scheduling blocks'') on March 20, July 4, and August 22, 2010.  We also made observations at 4136 MHz for the March and July sources, which would have provided polarization measurements at 4 frequencies. However, we ultimately flagged all 4.1 GHz data due to overwhelming  RFI. Table \ref{tbl2} presents a summary of the observations, which includes the date of observation, the duration of the sessions, the frequencies observed, the VLA array, the restoring beam used for each session, the number of scans per source per session, and the characteristic RMS noise level in the Q and U maps. The sources for the March and July sessions were the same, and we observed those sources at all three frequencies. The August session observed additional sources. This new set of sources was observed at 4.4GHz and 4.9GHz only. The intent was to observe these sources at 7.6 GHz as well, but the D array observing season ended before a 4$^{th}$ scheduling block was carried out. Table \ref{sources} lists all the sources with a project name in column 1, the RA and Dec (J2000) in columns 2 and 3, respectively, the galactic longitude and latitude in columns 4 and 5, the angular distance between the line of sight and a line of sight passing through the center of the Rosette, $\xi$, in column 6. The total Clean Flux at 4.9GHz is given in column 7, and in column 8, the number of frequencies observed for each source, where the number 3 corresponds to the set of frequencies of [4.4 GHz, 4.9 GHz, $\&$ 7.6 GHz] and the number 2 corresponds to the set [4.4GHz and 4.9GHz]. The range in frequency between 4.4 GHz and 7.6 GHz allows us to obtain RM values that are as low as a few tens of rad m$^{-2}$, given the errors in the polarization measurements (see Section 3.2 below). The shorter range between 4.4GHz and 4.9GHz allows for measurements of large RM values without being affected by the ``n$\pi$ ambiguity''.

\section{Data Reduction}
All data reduction was performed with the Common Astronomy Software Applications (CASA) data reduction package. The calibration procedure is similar to that used in our prior Faraday rotation projects with the VLA, such as \citet{whi2009} and \citet{min1996}. The procedure for reducing and calibrating the data was as follows.
\begin{enumerate}
\item We flagged out measurements corrupted by radio frequency interference (RFI).  For all sessions, some antennas were completely flagged because of corrupted or missing data. We also implemented position corrections for a number of antennas. As well as usual systematic flagging procedures (e.g. ``Quack''), we visually inspected the data in order to manually remove RFI and other problems. 
\item Calibration of the array, consisting of determination of the complex gains and instrumental polarization parameters (``D factors''), as well as the right-left phase difference for the entire array, was carried out following the online $\emph{EVLA Continuum Tutorial}$ and supplemented by the handbook for the CASA program.\footnote{\footnotesize{For further reference on data reduction, see the NRAO EVLA Tutorial ``EVLA Continuum Tutorial 3C391". [http$://$casaguides.nrao.edu$/$index.php$?$title$=$EVLA$_{}$Continuum$_{}$Tutorial$_{}$3C391]}}
\item Polarized images of the sources were made from the calibrated visibility data with the CASA task CLEAN. CLEAN is a task that Fourier transforms the data to form the ``dirty map'' and ``dirty beam'', carries out the CLEAN deconvolution algorithm, and restores the image by convolving the CLEAN components with the restoring beam. We produced CLEANed maps of the Stokes parameters I, Q, U, and V. Different weighting schemes in the (u, v) plane were used in the different sessions. The weighting was set to uniform for the March and July sources, but natural weighting was used for the August sources in order to obtain a better signal to noise ratio for the weaker sources observed in that session. The restoring beam for the March and July sources, across all frequency bands, was 12\farcs8. For the August sources, the restoring beam was 19\farcs6. The larger restoring beam in the August 2010 session is due to the use of natural rather than uniform weighting in the (u,v) plane. All maps presented utilized external calibration only. A single iteration of phase-only self calibration did not produce an improved signal-to-noise ratio for our maps.
\end{enumerate}

\subsection{Imaging the Sources}
Having obtained the maps of the Stokes parameters I, Q, U, and V for each source at each frequency, we generated maps of the linear polarized intensity, L, and the polarization position angle, $\chi$
\begin{equation}
L=\sqrt{Q^2+U^2}
\end{equation}
\begin{equation}
\chi=\frac{1}{2} \tan^{-1}({\frac{U}{Q}})
\end{equation}
For each source and frequency, we worked with images of I, L, and $\chi$. Examples of the images of two of our sources are shown in Figures \ref{figmaps15} and \ref{figmaps14}. Figure \ref{figmaps15} shows the I, L, and $\chi$ maps of a point source (to the D array), I15, that was found to have a large RM (633 $\pm$ 14 rad m$^{-2}$). Figure \ref{figmaps14} shows a source, O2, which is resolved to the D array and possesses structure. 

\subsection{Determination of Rotation Measures}
\begin{figure}[ht!]
\centering
\subfigure[4.4GHz]{
\includegraphics[scale=.3]{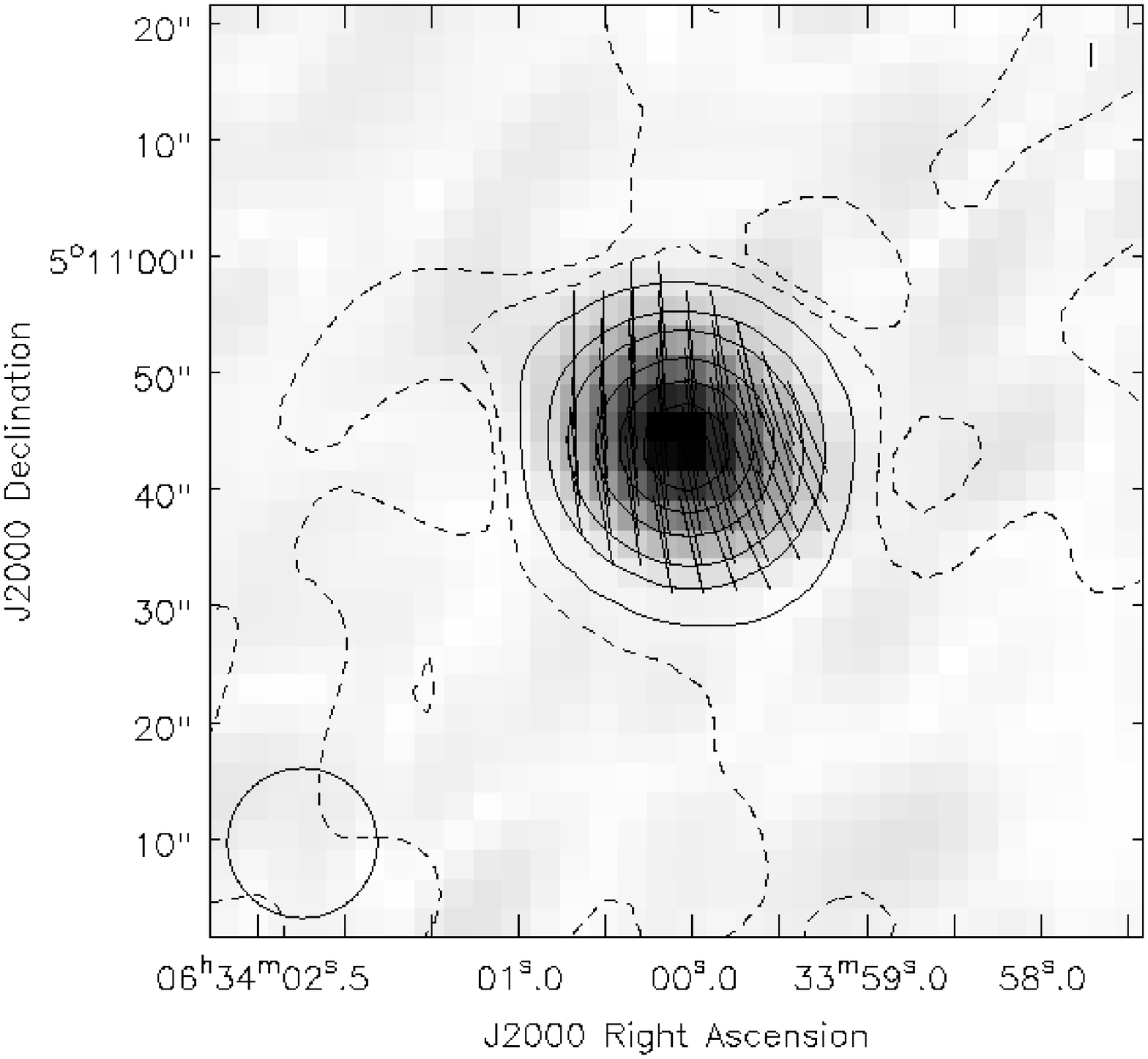}}
\subfigure[4.9GHz]{
\includegraphics[scale=.3]{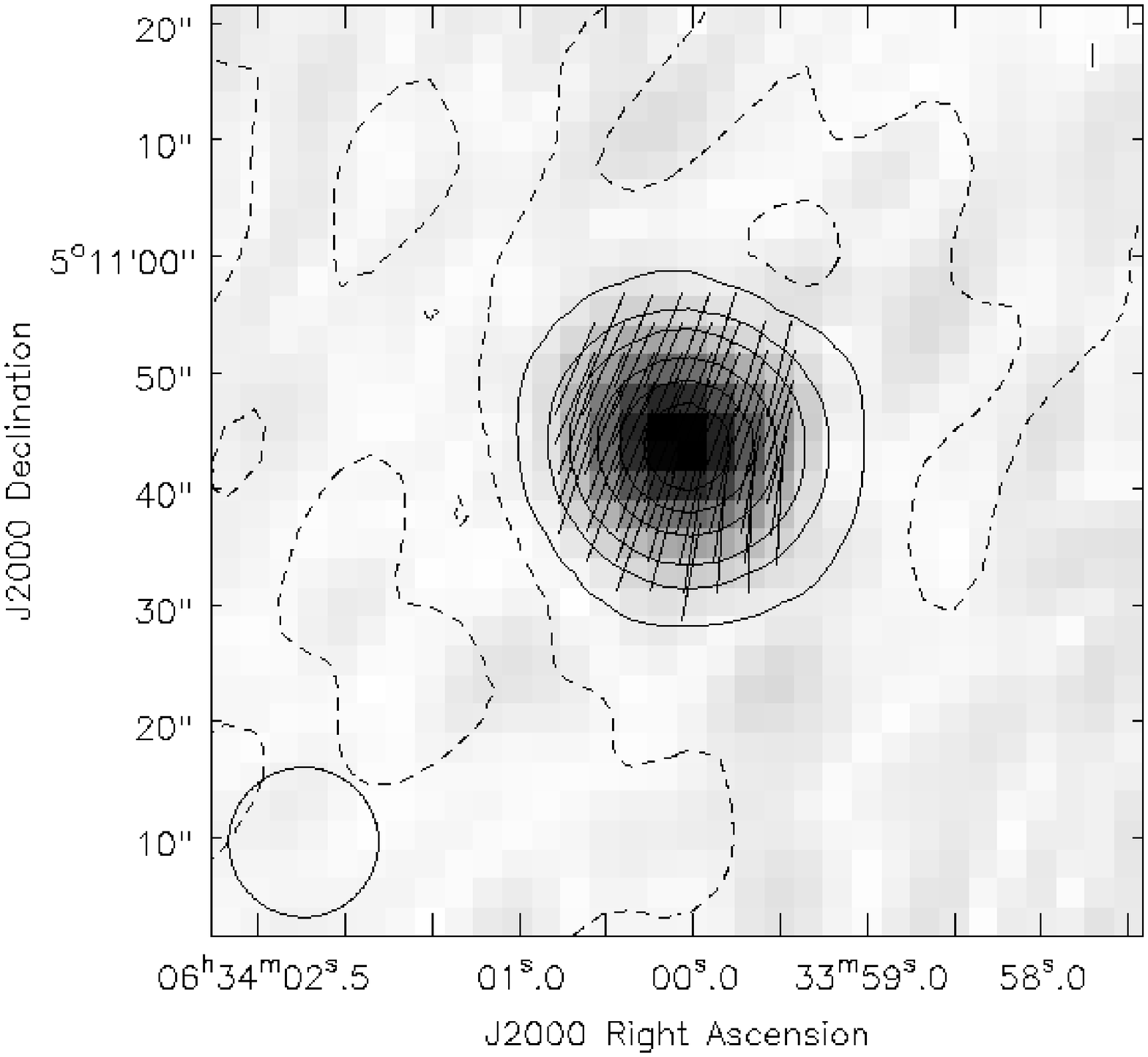}}
\subfigure[7.6 GHz]{
\includegraphics[scale=.3]{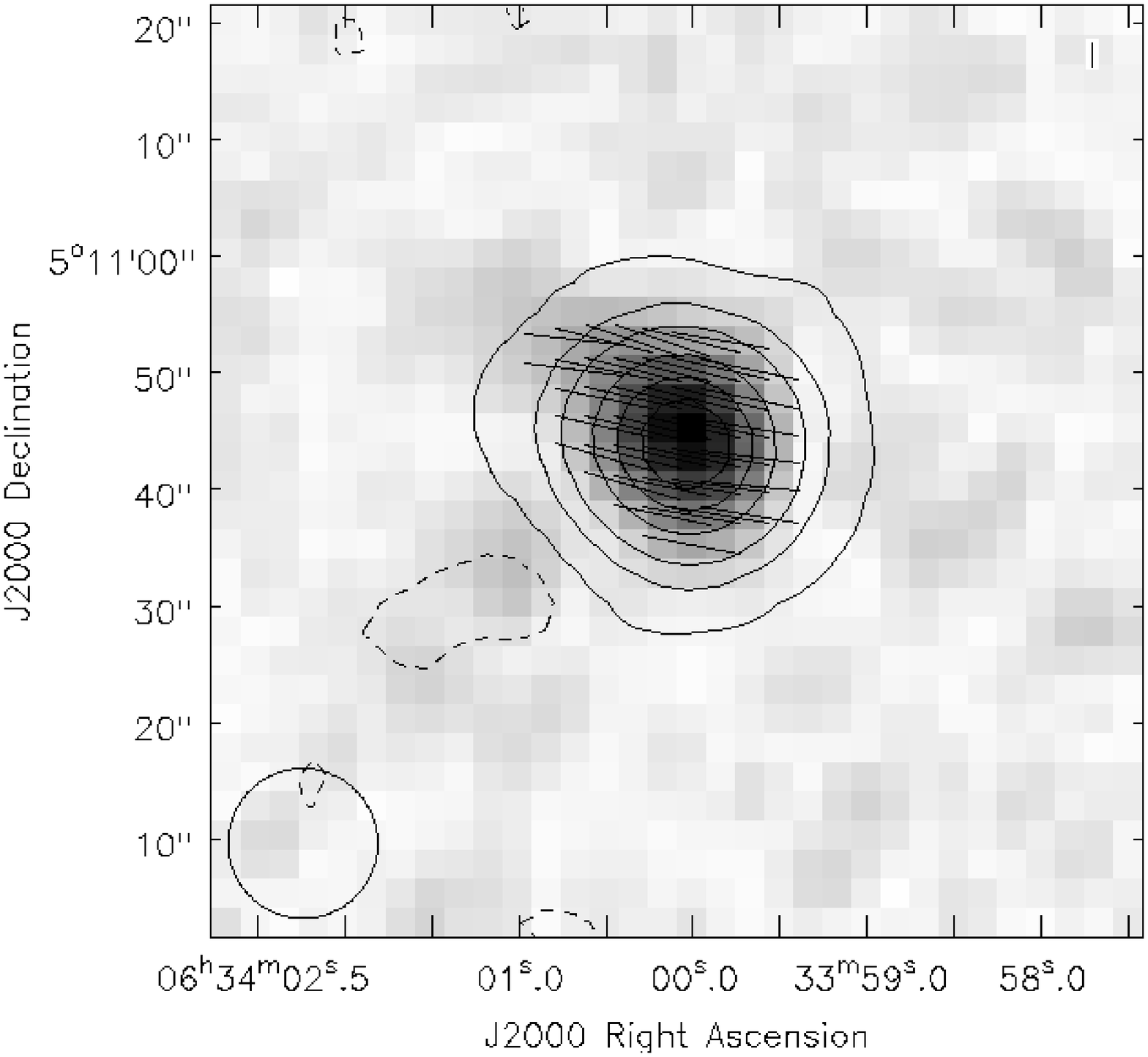}}
\caption{Maps of I15 at the three frequencies of observation. The vectors show the polarization position angle, $\chi$, the gray scale  is the polarized intensity, L, and the contours are the intensity, I,  with contour levels set to -2, -1, 2, 10 20 40 60 and 80$\%$ of the peak intensity, which is 19.0 mJy/beam at 4.4GHz. The dashed contours indicate negative intensities, illustrating the level of noise and map imperfections. The circle in the lower left corner indicates the restoring beam.}
\label{figmaps15}
\end{figure}

\begin{figure}[ht!]
\centering
\subfigure[4.4 GHz]{
\includegraphics[scale=.3]{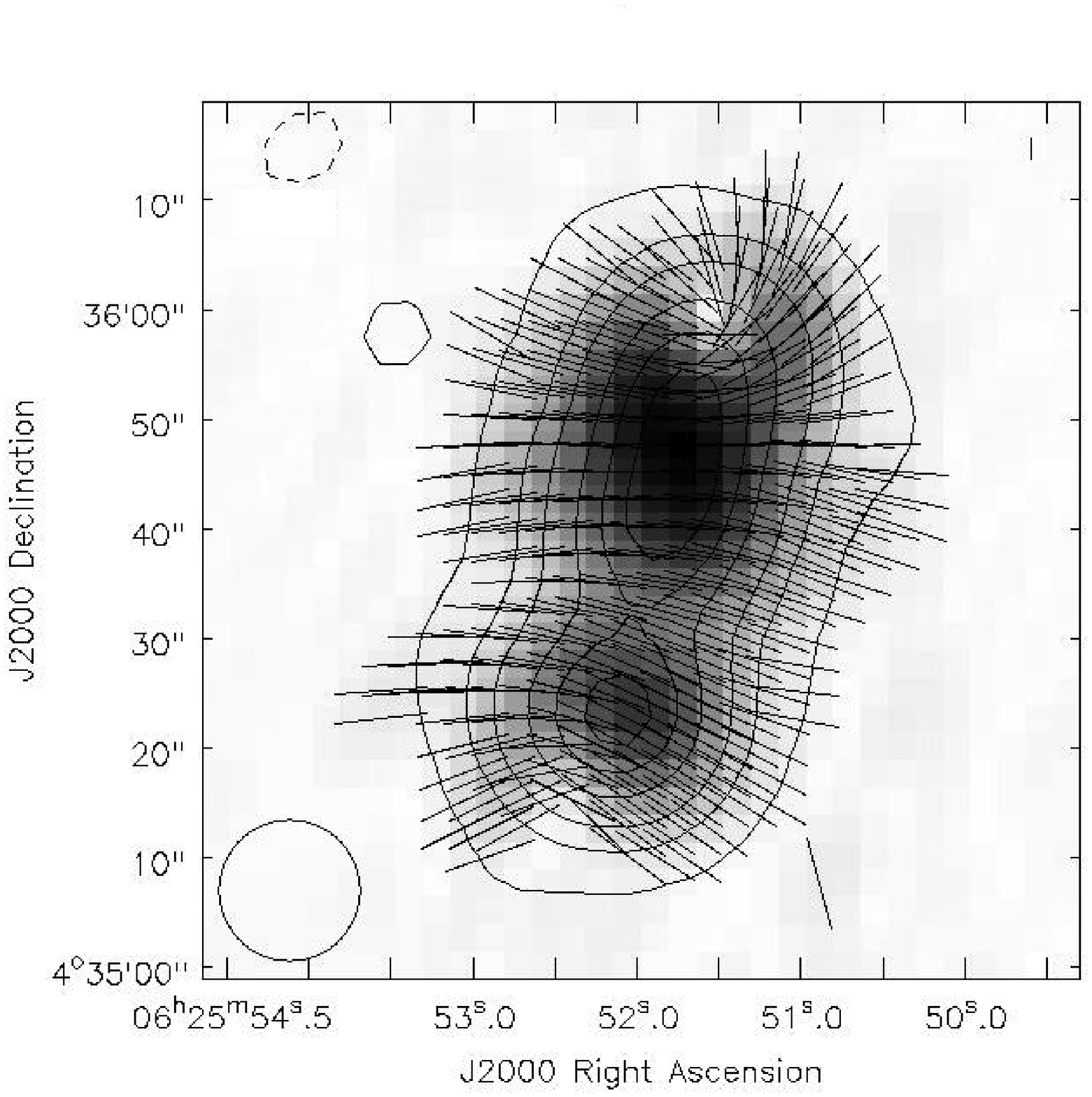}}
\subfigure[4.9 GHz]{
\includegraphics[scale=.3]{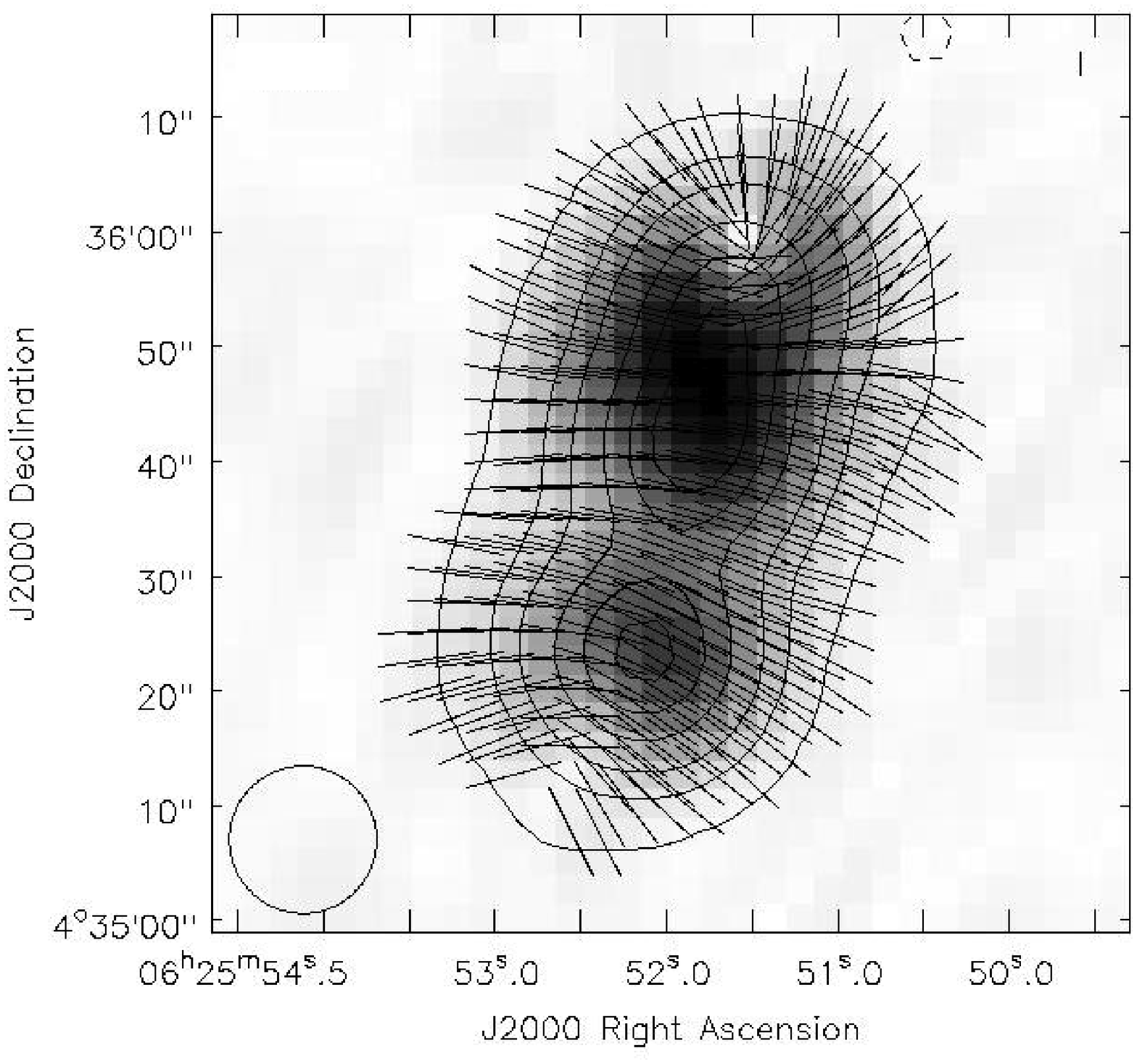}}
\subfigure[7.6 GHz]{
\includegraphics[scale=.3]{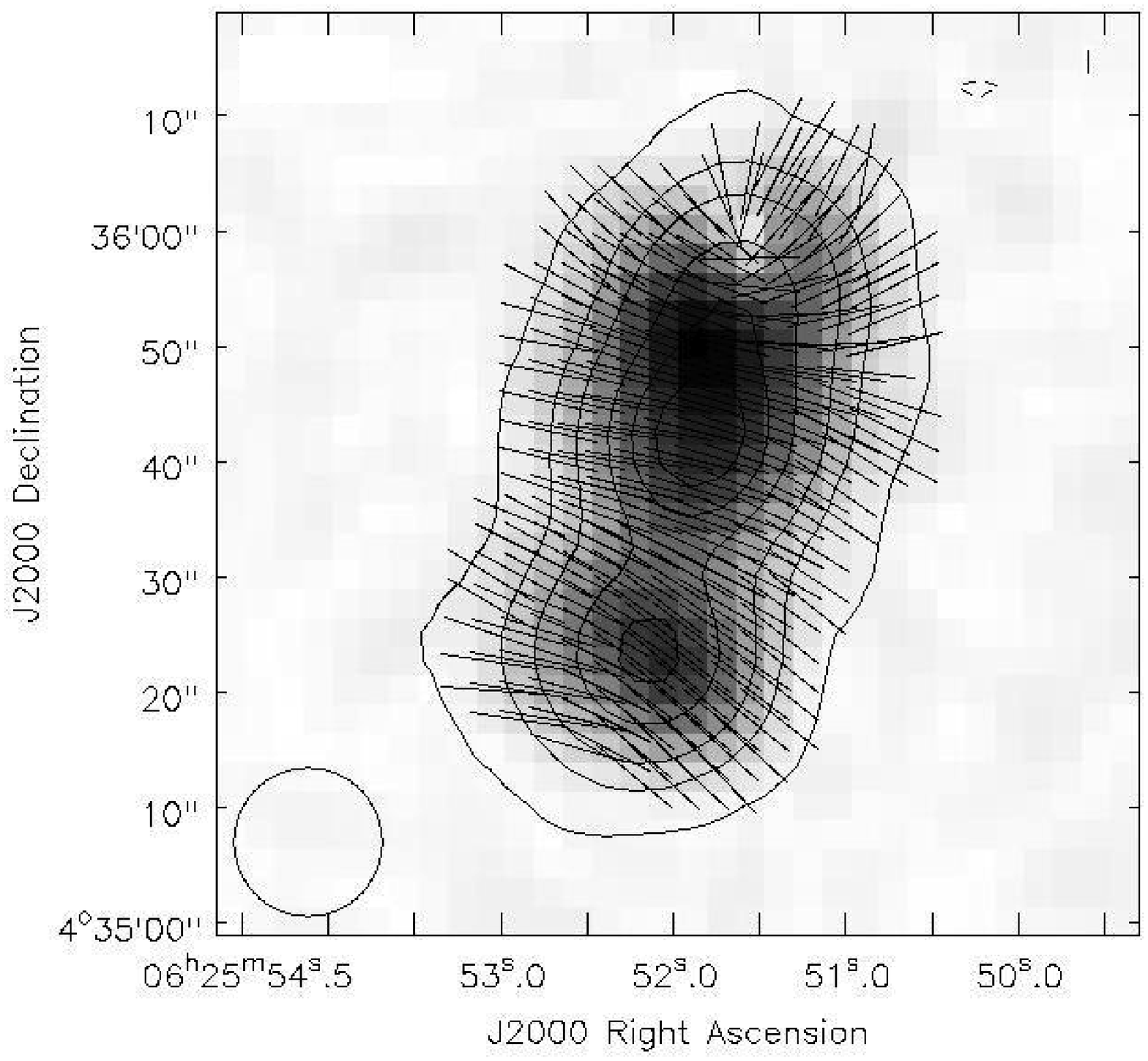}}
\caption{Maps of O2 in the same format as Figure \ref{figmaps15}. The contours of total intensity are at -2, -1, 2, 10 20 40 60 and 80$\%$ of the peak intensity, which is 10.2 mJy/beam at 4.4GHz.}
\label{figmaps14}
\end{figure}

\begin{figure}[hbt!]
\begin{center}
\includegraphics[scale=0.45]{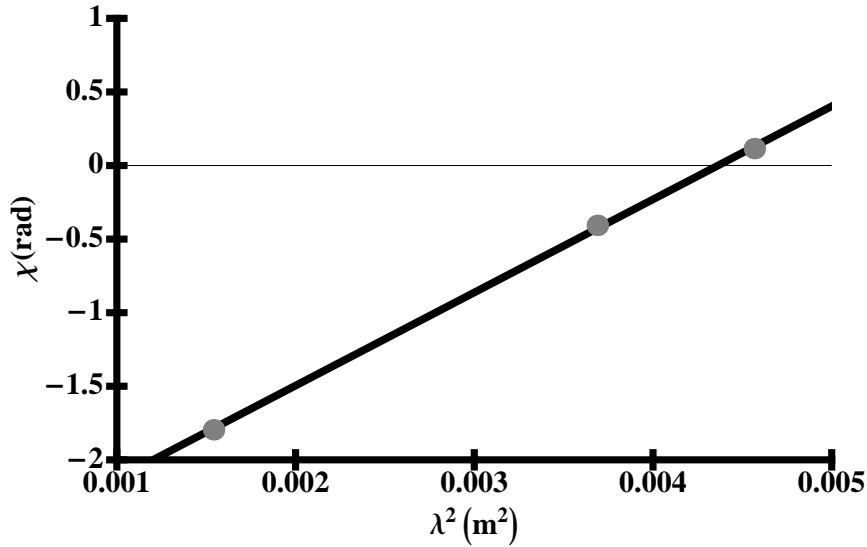}
\caption{A plot of the polarization position angle $\chi$ in radians against the square of the wavelength in [m$^2$] for the interior source I15 (Image shown in Figure \ref{figmaps15}). The value of the fit RM is 633 $\pm$ 14 rad m$^{-2}$. Error bars are contained within the plotted symbols.}
\label{fig15}
\end{center}
\end{figure}

\begin{figure}[htb!]
\begin{center}
\includegraphics[scale=0.65]{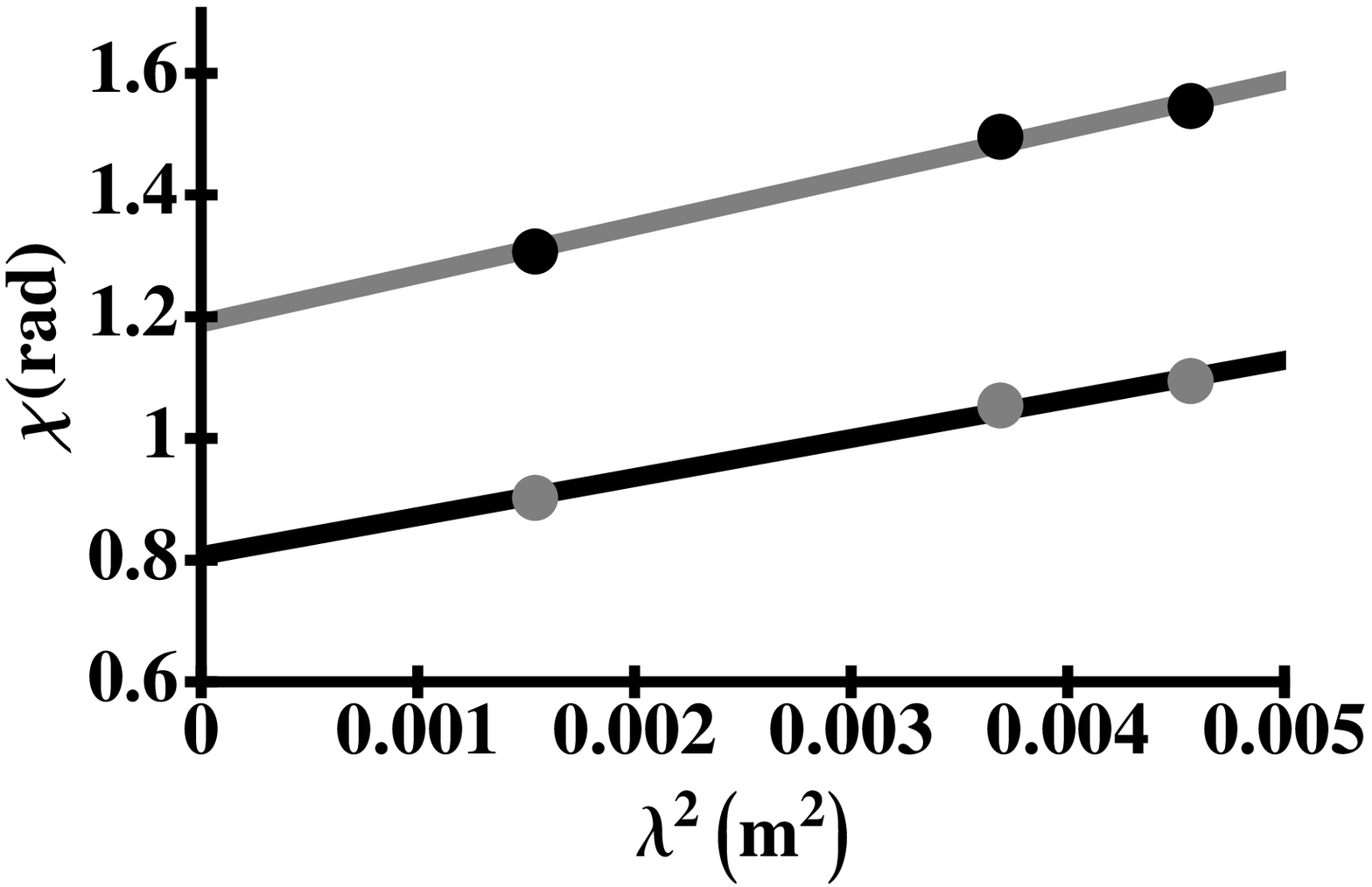}
\caption{Polarization position angle data for source O2, in the same format as Figure \ref{fig15}. The two sets of data points present measurements for the two components of the source seen in Figure \ref{figmaps14}. The solid black points represent the data for the north component (component (a) in Table 3), and the solid gray points are for the south component (component (b) in Table 3). The fit RM values are 80 $\pm$ 8 rad m$^{-2}$ for component (a) and RM= 64 $\pm$ 6 rad m$^{-2}$ for component (b). Error bars are contained within the plotted symbols.}
\label{figI14}
\end{center}
\end{figure}

In this section, we describe how we obtained RMs from data of the sort shown in Figures \ref{figmaps15} and \ref{figmaps14}. We first identified a local maximum in the polarized intensity in the 4.4GHz map. We then measured the polarization position angle $\chi$ at this location for the 2 or 3 frequencies available for this source. Since the sources from the August scheduling block have only two data points, the RM was calculated from Equation (\ref{RM}). There are also larger errors associated with the sources from the August scheduling block due to having only two data points that are only slightly separated in frequency. There are three data points for the sources from the March and July scheduling blocks and the RM was calculated by plotting the polarization position angle, $\chi$, against $\lambda^{2}$ and fitting a line to this relationship. An example of this is illustrated in Figure \ref{fig15}, for the source I15, and Figure \ref{figI14} for the source O2.  All of our RM values were positive except for I18 and a component of O14.  Two of the sources, I3 and O11, were depolarized, so we did not obtain a RM for them.

The source I2 requires additional comments. I2 is an interior source from the March and July scheduling blocks. Usually, this would mean that we had polarization data at three frequencies. However, the 4.4GHz data for I2  were excluded from the calculation of the RM measurement. The 4.4GHz data failed a test for data quality that we applied to our observations, as follows. For each source and frequency, the data from each scan was mapped. As discussed in Section 2 above, each source was typically observed for 5 scans during the 6 hour observing session. These scan maps were made in all polarization parameters as well as the total intensity I. The purpose of this exercise was to make sure that no systematic changes occurred during the observing session, due to incorrect correction for instrumental polarization, or similar effects. Once it was determined that the polarization data were ``stationary'' during the observing session, and that no drastically flawed data were present, I, Q, and U maps as well as maps of the derived quantities L and $\chi$, were made with all available data. Unlike the other sources for which values of $\chi$ were within 1 $\sigma$ of the mean values, the $\chi$  time series for I2 showed scan-to-scan variations larger than noise. We examined I2 at 4.9GHz and 7.6 GHz, and determined that inconsistent polarization position angles were not present at the higher frequencies. Since we only used two frequencies in determining the RM value for I2, there is a larger error associated with this source. The degree of linear polarization for I2 was extremely low, 0.1$\%$ at 4.4GHz, and we attribute the variations of $\chi$ to residual instrumental polarization artifacts, which can appear with low values of the degree of linear polarization \citep{sak1994}. We retain I2 as one of the sources in our sample because we believe the data from the two higher frequencies are adequate to determine $RM$.  The $RM$ for I2 is consistent with the values for adjacent sources that we determined from measurements at three frequencies, indicating that n-$\pi$ ambiguities are not a problem. 

The fits to the data shown in Figures \ref{fig15} and \ref{figI14} are sufficiently good to give us confidence that we have an accurate measure of the RM. Nonetheless, there can be a residual concern that the RM is larger than the value resulting from our fit, and that there is one or more rotations of $\pi$ radians between the  frequencies observed. To exclude this possibility and demonstrate that the 3 frequency RM fits were accurate, we made a $\chi$($\lambda^{2}$) fit within the 4.4GHz bandpass for those sources with RM $\geq$ 500 rad m$^2$. As described above, the 4.4GHz spectral window had 64 channels of 2MHz bandwidth. The end channels on both ends of the bandpass were discarded, and the remaining channels averaged to 5 sub-IF channels of 22 MHz each. A fit of $\chi$($\lambda^{2}$) = $\chi_0$ + RM$\lambda^2$ was then redone over the 4.4GHz spectral window. In all cases, the RM from this procedure agreed, within the errors, with the values obtained by fitting to two or three frequencies of 4.4, 4.9, and 7.6 GHz.  A final check of the data set was to examine the degree of linear polarization,
\begin{displaymath}
 m= \frac{L}{I}, 
\end{displaymath}
where I is the total intensity, for each source or source component at each of the frequencies of observation. If the degree of linear polarization is constant, this indicates that the Faraday rotation occurs in an external medium, such as the Galactic interstellar medium. A case where m is a function of frequency, with a smaller $m$ at lower frequencies, indicates internal Faraday rotation and depolarization within the synchrotron emitting source. The dependence of $\chi$ on $\lambda$ is then not proportional to $\lambda^{2}$, and a fit of the type we have done could yield an inaccurate estimate of the RM. 

For each source or source component with measurements at 3 frequencies, the weighted mean degree of linear polarization $\bar{m}$ was calculated from the measurements of $m$ at each of the frequencies.  The weighting was with the error on $m$, calculated from the noise level in the $Q$ and $U$ maps.  We then calculated the reduced $\chi_{\nu}^2$ statistic for the 3 measurements about this mean (with $\nu = 2$ degrees of freedom), and chose as a flag threshold a value of $\chi_{\nu}^2 = 3.9$, which corresponds to a 2 \% probability of constancy of $m$ with frequency, for three measurements \citep{bev1969}.  

We considered the $\chi_{\nu}^2$ statistic as a screening operation rather than a definitive test, since the error in $m$ was calculated from the $Q$ and $U$ noise levels on blank portions of the image; such a procedure can underestimate the true error on a portion of the source where $L$ or $I$ is large.  Of the 16 sources or source components (excluding I2) with observations at three frequencies, 9 passed this screening operation for $m$ being independent of frequency, and thus unaffected by depolarization. 

We then carefully examined the data for the remaining sources in more detail. We found that in nearly every case, depolarization could be excluded, and we concluded that the blank field noise measurements underestimate the true errors in $m$. For example, for 2 source components (O2a and O5b) the excessive $\chi_{\nu}^2$ was due to $m$ at 7.6 GHz being slightly lower than at 4.4 and 4.9 GHz.  This is the opposite of the behavior for Faraday depolarization, and shows that our measurements are not affected by depolarization. 

In 4 of the 5 remaining cases, the decrease in $m$ from 7.6 to 4.9 GHz was very small (i.e. $\leq 13$ \%), and we believe the high values of $\chi_{\nu}^2$ are due to the low estimate of measurement errors on $m$.  In all of the aforementioned cases, we feel that a fit of $\chi(\lambda^2)$ gives a good estimate of the $RM$ due to the Galactic ISM, unaffected by the depolarization within the source.  A point in support of this contention is the fact that 3 of these components were in double sources, and the $RM$s of the two components were in satisfactory agreement (see Table 3, described below).  

The only source for which depolarization might be present is I14b.  It was flagged by the $\chi_{\nu}^2$ screening criterion, and the degrees of linear polarization at 4.4, 4.9, and 7.6 GHz are $0.018 \pm 0.001$, $0.023 \pm 0.001$, and $0.031 \pm 0.001$, respectively. These measurements seem to show a progression in $m$ with increasing frequency, as well as a reduced $\chi_{\nu}^2$ value formally inconsistent with constancy.  These data may indicate depolarization, in which case a source-associated rotation of the position angle, independent of the Galactic ISM, might occur.  Furthermore, in this case there is a difference in $RM$ between the two components of the source (see Table 3 below), although a linear fit to the $\chi$ versus $\lambda^2$ data was obtained. Although this difference in $RM$ between two source components with a small angular separation could indicate a problem with depolarization, it could also be an interesting probe of small scale variations in the nebula, as discussed in Section 4.2 below. In the remainder of this paper, we will use the data for component I14b, with the recognition that the inferred RM might contain a component due to the source itself rather than the Galactic ISM. 

A similar test was undertaken, with a corresponding reduction in the degrees of freedom, for the 12 sources or source components with observations at two frequencies, Only 1 source (O7) had a $\chi_{\nu}^2$ for 1 degree of freedom that exceeded the 2 \% probability threshold and therefore merited closer examination.  We concluded that the large $\chi_{\nu}^2$ was due to small inferred errors on the $m$ values at the 2 frequencies; the $m$ values at 4.4 and 4.9 GHz are in good agreement, with $m_{4.4} > m_{4.9}$.  Internal Faraday depolarization or depolarization by a plasma screen in front of the source cannot be occurring in this case.   

To conclude, with the probable exception of I2, and the possible but not certain case of I14b, all of the $RM$ values obtained from our sources and source components appear to be measures of the Galactic ISM.  

Our results on the polarization properties of our sources and the resultant RM values are shown in Table 3. The first column of Table \ref{RMvalues} lists the source name. Duplication of sources in this column indicates that there were two components to the source for which we were able to obtain RM values. Each source has two or three associated rows in the table, and subsequent components of the same source also have two or three rows. These rows give data for the two or three frequencies of observation. Column 2 identifies the components of the duplicated source as either (a) or (b). There were 9 sources that had two components. Column 3 lists the frequency associated with the data for the source, column 4 lists the linear polarized intensity, L (mJy/beam), and the associated error, column 5 the degree of linear polarization, m. Column 6 is the polarization position angle $\chi$ and the associated error,  and column 7 has the RM with associated errors. Since the RM is obtained by fitting a line to the $\chi$($\lambda^2$) data for the March and July sources, and by Equation (\ref{RM}) for the August sources, column 7 has one value per source component.

\subsection{Comparison of RM Measurements with \citet{tay2009}}
\begin{figure}[ht!]
\begin{center}
\includegraphics[scale=0.55]{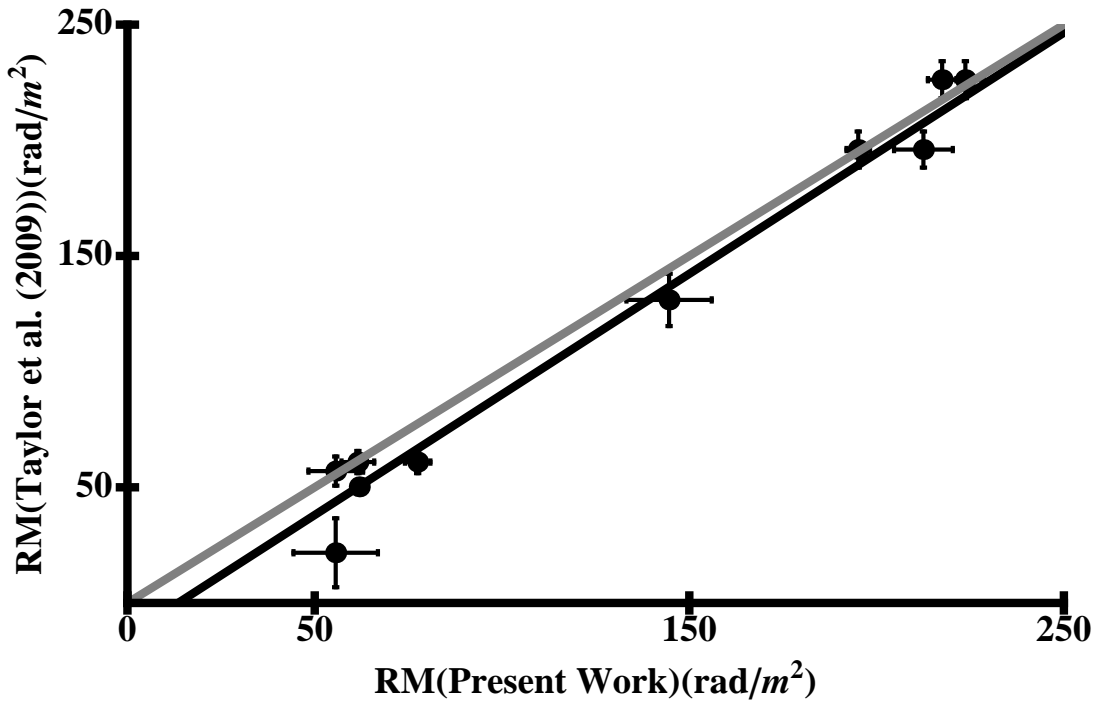}
\caption{A comparison of RM values from \citet{tay2009} and the present study. The lighter solid line represents the case of perfect agreement, and the heavy solid line represents a weighted least-squares fit to the data.}
\label{figtaylor}
\end{center}
\end{figure}
\citet{tay2009} re-analyzed data from the NRAO VLA Sky Survey (NVSS) in order to obtain RMs for 37,543 radio sources. That study provided RMs for the sky north of -40$^\circ$ in declination. We can compare our RM values with the previously derived RM values by \citet{tay2009} for seven of our sources in common with \citet{tay2009}. There are two reasons for carrying out this comparison. First, it serves as a check on our data and method of data analysis. Second, there are inconsistent reports in the literature regarding the accuracy of the \citet{tay2009} results. In a study of the magnetic field in the direction of the Galactic poles, \citet{mao2010} found discrepancies between their RM measurements and those of \citet{tay2009}. A comparison of RM values from \citet{tay2009} and independent measurements was also made by \citet{van2011}, with the VLA. \citet{van2011} found generally satisfactory agreement, although there was a population of outliers as well as an apparent systematic bias at RM $\simeq$ 50 - 100 rad m$^{-2}$ (see Figure 4 of \citet{van2011}). A full discussion of the comparison between independent RM measurements and the RM values from \citet{tay2009} is given in Section 4.2 of \citet{har2011}. We think it worthwhile to make additional comparisons between \citet{tay2009} and independent measurements made specifically for the purpose of measuring Faraday rotation. Figure \ref{figtaylor} illustrates the comparison  of our RM measurements with those of \citet{tay2009}. The sources that are in common are all exterior sources, O1, O2, O4, O7, O9, O15, $\&$ O17. None of our interior sources were contained in the catalog of \citet{tay2009}. Our RM values compare favorably with those of \citet{tay2009}. The light solid line in Figure \ref{figtaylor} shows the case of perfect agreement between the two sets of measurements, and this is clearly a satisfactory representation of the data. A weighted least squares linear fit to the data shown in Figure \ref{figtaylor} gives a slope of m= 1.04 $\pm$ 0.04  and an intercept of b= -14.1 $\pm$ 8.1 rad m$^{-2}$ (heavy solid line). The good agreement between the two sets of measurements is consistent with the assessment of \cite{van2011}, and gives confidence in our $RM$ measurements. We note that this does not address the question of the systematic error in some of \cite{tay2009} $RM$s that was pointed out by \cite{van2011}.

\section{Observational Results and Modeling in Terms of the Interaction of an HII Region with the Interstellar Medium}
\begin{figure}[!ht]
\begin{center}
\includegraphics[scale=0.5]{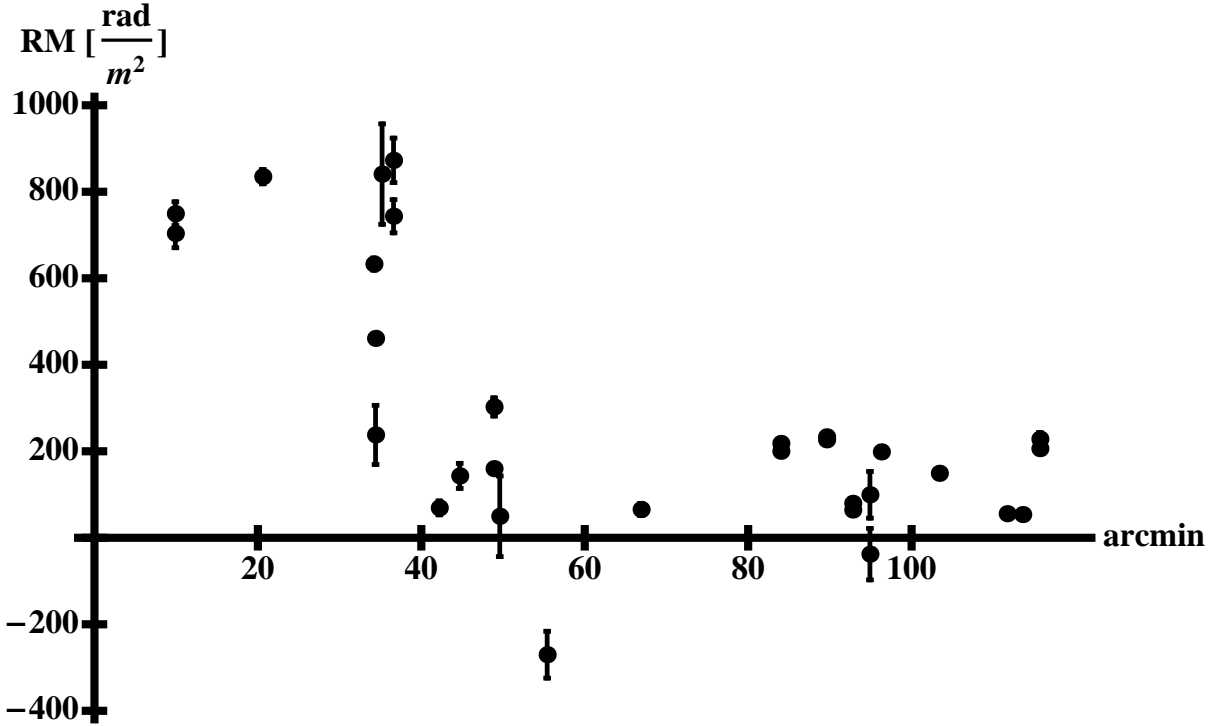}
\caption{The RM (rad m$^{-2}$) for each source and source component in Table 3 versus angular distance (arcminutes) from the center of the Rosette Nebula. All the sources, and components, are represented on the graph along with the associated error bars.}
\label{figrmarc}
\end{center}
\end{figure}

The first question in the analysis is whether the RM data from Table \ref{RMvalues} show evidence for an RM enhancement associated with the Rosette Nebula. Such an enhancement is  illustrated and clearly seen in Figure \ref{figrmarc}. In Figure \ref{figrmarc}, we plot the measured RM versus angular distance from the center of the Rosette Nebula, which we take to be the center of the NGC 2244 star cluster as given by \citet{ber2002} (see Introduction). A very clear signature of a Faraday rotation enhancement is seen for the 6 lines of sight (9 sources and source components) with angular separation $\leq$ 40 arcminutes from the nebular center. The excess $RM$ due to the Rosette Nebula is also visible in Figure 1, in which the size of the plotted symbol for each source is dependent on $RM$.  Those sources viewed through the Rosette have larger $RM$s.  The mean RM for sources seen through the Rosette Nebula is 675 rad m$^{-2}$, with a range of 200 $\leq$ RM $\leq$ 900 rad m$^{-2}$. Lines of sight that are more than 40 arcminutes from the center of the nebula have a mean of 147 rad m$^{-2}$, with a standard deviation of 77 rad m$^{-2}$. In calculating the mean and standard deviation of the background, we have excluded the two sources in our sample with a negative RM (I18, RM=-270 $\pm$ 54 rad m$^{-2}$ $\&$ O14(b), RM=-38 $\pm$ 60 rad m$^{-2}$). It is unclear whether the negative RMs have Galactic or extragalactic origins. The RM in both cases was obtained from measurements at only 2 frequencies. As presented in Table \ref{RMvalues}, both the polarized intensity and degree of linear polarization for those 2 sources are low. Although we include these sources in Table \ref{RMvalues} because they passed our selection criteria, we do not include them in our calculation of the Galactic mean background. We interpret this mean background as due to the Galactic Faraday rotation in this part of the sky, which is independent of the Rosette Nebula. The data in Figure \ref{figrmarc} show a ``RM anomaly'' of 50-750 rad m$^{-2}$ associated with the Rosette Nebula. This is comparable to, and perhaps slightly smaller than that reported for the Cygnus OB1 association by \citet{whi2009}. However, the \citet{whi2009} result is more ambiguous because of the angular proximity of other HII regions as well as other Galactic objects, which confuse measurements in that field.
\clearpage
\begin{deluxetable}{ccccccc}
\tablecaption{Polarization Results \label{RMvalues}}
\tablehead{
\colhead{Source} & \colhead{Component} & \colhead{$\nu$ [GHz]} & \colhead{L [mJy/beam]} & \colhead{m [$\%$]} & 
\colhead{$\chi$ [$^o$]}  & \colhead{RM [rad m$^{-2}$]}}
\startdata
I1 &  & 4.4 & 0.25 $\pm$ 0.03 & 2 & -24.5 $\pm$ 3.5 & {\multirow{2}{*}{50 $\pm$ 93}} \\ 
&  & 4.9 & 0.26 $\pm$ 0.03 & 3 & -27.0 $\pm$ 3.4 & \\
\hline
I2 & & 4.4 & 0.15 $\pm$ 0.03 & 0.1 & 6.8 $\pm$ 5 & {\multirow{3}{*}{69 $\pm$ 16}} \\
&  & 4.9 & 0.37 $\pm$ 0.03 & 0.2 & -14.8 $\pm$ 1.9 & \\
&  & 7.6 & 1.70 $\pm$ 0.04 & 1.0 & -23.3  $\pm$ 0.7 &\\
\hline
I6 & a  & 4.4 & 0.90 $\pm$ 0.04 & 8 & 13.4 $\pm$ 1.1 & {\multirow{2}{*}{743 $\pm$ 38}} \\
& a & 4.9 & 0.78 $\pm$ 0.04 & 7 & -23.7 $\pm$ 1.7 & \\
\hline
I6 & b & 4.4 & 0.60 $\pm$ 0.03 & 9 & -4.7 $\pm$ 1.6 & {\multirow{2}{*}{873 $\pm$ 51}} \\
& b & 4.9 & 0.62 $\pm$ 0.03 & 10 & -48.6 $\pm$ 2.1 & \\
\hline
I7 & a & 4.4 & 0.93 $\pm$ 0.03 & 10 & -45.4  $\pm$ 1.0 &  {\multirow{2}{*}{749$\pm$27}} \\
& a & 4.9 & 0.87 $\pm$ 0.03 & 10 & -83.1 $\pm$ 1.0 & \\
\hline
I7 & b & 4.4 & 0.79 $\pm$ 0.03 & 5 & 0.02 $\pm$ 1.15 & {\multirow{2}{*}{704$\pm$33}} \\
& b & 4.9 & 0.69 $\pm$ 0.03 & 4 & -35.4 $\pm$ 1.3 & \\
\hline
I8 & a & 4.4 & 1.11 $\pm$ 0.04 & 14 & 46.6 $\pm$ 1.0 & {\multirow{3}{*}{461$\pm$4}} \\
& a & 4.9 & 0.99 $\pm$ 0.03 & 13 & 22.9 $\pm$ 0.9 & \\
& a & 7.6 & 0.64 $\pm$ 0.04 & 14 & -33.4 $\pm$ 1.5 & \\
\hline
I8 & b & 4.4 & 0.41 $\pm$ 0.04 & 7 & 28.8 $\pm$ 2.6 & {\multirow{3}{*}{238$\pm$ 73}} \\
& b & 4.9 & 0.28 $\pm$ 0.03 & 6 & 6.1 $\pm$ 3.2 & \\
& b & 7.6 & 0.21 $\pm$ 0.03 & 7 & -13.5 $\pm$ 4.7 & \\
\hline
I10 &  & 4.4 & 0.23 $\pm$ 0.04 & 2 & 108.2 $\pm$ 5.1 &  {\multirow{2}{*}{841 $\pm$117}} \\
&  & 4.9 & 0.29 $\pm$ 0.03 & 3 & 65.9 $\pm$ 3.3 & \\
\hline
I12 &  & 4.4 & 1.3 $\pm$ 0.04 & 5 & 61.5 $\pm$ 1 & {\multirow{3}{*}{ 835$\pm$17}} \\
&  & 4.9 & 1.25 $\pm$ 0.04 & 5 & 21.9 $\pm$ 0.7 & \\
&  & 7.6  & 1.02 $\pm$ 0.05& 5 & -82.7 $\pm$ 1.3 & \\
\hline
I14 & a & 4.4 & 3.30 $\pm$ 0.04 & 25 & 59.1  $\pm$ 0.3 & {\multirow{3}{*}{160 $\pm$3}} \\
& a & 4.9 & 2.96 $\pm$ 0.03 & 26 & 51.4 $\pm$ 0.3 & \\
& a & 7.6 & 1.78 $\pm$ 0.04 & 25 & 31.4 $\pm$ 0.6 & \\
\hline
I14 & b & 4.4 & 0.75 $\pm$ 0.04 & 2 & 106.4  $\pm$ 1.3 & {\multirow{3}{*}{303$\pm$21}} \\
& b & 4.9 & 0.90 $\pm$ 0.03 & 2 & 87.4 $\pm$ 1.0 & \\
& b & 7.6 & 0.86 $\pm$ 0.04 & 3 & 53.0 $\pm$ 1.3 &  \\
\hline
I15 &  & 4.4 & 0.93 $\pm$ 0.04 & 5 & 6.6  $\pm$ 1.1 & {\multirow{3}{*}{633$\pm$14}} \\
&  & 4.9 & 0.89 $\pm$ 0.04 & 5 & -23.3 $\pm$ 1.2 & \\
&  & 7.6 & 0.56 $\pm$ 0.04 & 4 & -102.9 $\pm$ 1.9 & \\
\hline
I16 &  & 4.4 & 0.94 $\pm$ 0.03  & 11 & 84.2  $\pm$ 1.0 & {\multirow{2}{*}{143$\pm$29 }} \\
&  & 4.9 & 0.81 $\pm$ 0.03 & 10 & 77.0 $\pm$ 1.1 & \\
\hline
I18 &  & 4.4 & 0.45 $\pm$ 0.03 & 4 & 28.2 $\pm$ 1.6 &  {\multirow{3}{*}{-270$\pm$54}} \\
&  & 4.9& 0.38 $\pm$ 0.03 & 4 & 41.8 $\pm$ 2.3 & \\
\hline
O1 &  & 4.4 & 5.75 $\pm$ 0.05 & 6 & -55.5 $\pm$ 0.2 & {\multirow{2}{*}{54$\pm$7}} \\
&  & 4.9 & 5.09 $\pm$ 0.06 & 6 & -58.2 $\pm$ 0.3 & \\
\hline
O2 & a & 4.4 & 4.89 $\pm$ 0.05 & 5 & 88.6  $\pm$ 0.3 & {\multirow{3}{*}{80$\pm$8}} \\
& a & 4.9 & 4.42 $\pm$ 0.04 & 5 & 85.7  $\pm$ 0.3 & \\
& a & 7.6 & 3.16 $\pm$ 0.06 & 4 & 74.9 $\pm$ 0.5 & \\
\hline
O2 & b & 4.4 & 3.43 $\pm$ 0.05 & 4 & 62.7 $\pm$ 0.4 & {\multirow{3}{*}{64$\pm$6}} \\
& b & 4.9 & 3.20 $\pm$ 0.05 & 4 & 60.4  $\pm$ 0.5 & \\
& b & 7.6 & 2.59 $\pm$ 0.06 & 4 & 51.7 $\pm$ 0.7 &  \\
\hline
O4 & a & 4.4 & 4.28 $\pm$ 0.04 & 23 & 49.7 $\pm$ 0.3 & {\multirow{3}{*}{200$\pm$1}} \\
& a & 4.9 & 3.92 $\pm$ 0.04 & 23 & 39.5 $\pm$ 0.3 & \\
& a & 7.6 & 2.69 $\pm$ 0.04  & 23 & 15.0 $\pm$  0.5 & \\
\hline
O4 & b & 4.4 & 1.60 $\pm$ 0.05 & 6 & -9.4 $\pm$ 0.8 & {\multirow{3}{*}{ 218 $\pm$8}} \\
& b & 4.9 & 1.45 $\pm$ 0.05 & 6 & -19.1 $\pm$ 0.8 & \\
& b & 7.6 & 1.10 $\pm$ 0.04 & 6 & -47.0 $\pm$ 1.1 & \\
\hline
O5 & a & 4.4 & 1.24 $\pm$ 0.05 & 11 & 85.8 $\pm$ 1.1 &  {\multirow{3}{*}{206$\pm$10}} \\
& a & 4.9 & 1.05 $\pm$ 0.05 & 11 & 74.0 $\pm$ 1.3 & \\
& b & 7.6 & 0.73 $\pm$ 0.04 & 11 & 49.9 $\pm$ 1.8 & \\
\hline
O5 & b & 4.4 & 1.07 $\pm$ 0.05 & 4 & 79.0 $\pm$ 1.3 &  {\multirow{3}{*}{228$\pm$15}} \\
& b & 4.9 & 0.96 $\pm$ 0.05 & 4 & 69.6 $\pm$ 1.4 & \\
& b & 7.6 & 0.54 $\pm$ 0.04 & 3 & 39.6 $\pm$ 2.5 & \\
\hline
O7 &  & 4.4 & 4.03 $\pm$ 0.05 & 2 & 19.3 $\pm$ 0.4 & {\multirow{2}{*}{55$\pm$11}} \\
&  & 4.9 & 3.84 $\pm$ 0.06 & 2 & 16.5 $\pm$ 0.4 & \\
\hline
O9 & a & 4.4 & 4.00 $\pm$ 0.04 & 14 & 93.5 $\pm$ 0.3 & {\multirow{3}{*}{226$\pm$1}} \\
& a & 4.9 & 3.69 $\pm$ 0.05 & 14 & 81.9 $\pm$ 0.4 & \\
& a & 7.6 & 2.42 $\pm$ 0.05 & 14 & 54.2 $\pm$ 0.6 & \\
\hline
O9 & b & 4.4 & 2.93 $\pm$ 0.04 & 9 & 83.0 $\pm$ 0.4 & {\multirow{3}{*}{233$\pm$1}} \\
& b & 4.9 & 2.76 $\pm$ 0.05 & 10 & 71.0 $\pm$ 0.5 & \\
& b & 7.6 & 1.82 $\pm$ 0.05  & 10 & 42.5 $\pm$ 0.8 & \\
\hline
O14 & a & 4.4 & 0.55 $\pm$ 0.04 & 9 & -81.3 $\pm$ 2.0 & {\multirow{2}{*}{99$\pm$54}} \\
& a & 4.9 & 0.51 $\pm$ 0.04 & 9 & -86.3 $\pm$ 2.0 & \\
\hline
O14 & b & 4.4 & 0.49 $\pm$ 0.04 & 9 & 9.7 $\pm$ 2.15 & {\multirow{2}{*}{-38$\pm$60}} \\
& b & 4.9 & 0.46 $\pm$ 0.04 & 9 & 11.6 $\pm$ 2.2 & \\
\hline
O15 &  & 4.4 & 22.67 $\pm$ 0.06 & 8 & 18.3 $\pm$ 0.1 & {\multirow{3}{*}{65$\pm$14}} \\
&  & 4.9 & 20.80 $\pm$ 0.06 & 8 & 16.4  $\pm$ 0.1 & \\
&  & 7.6 & 14.87 $\pm$ 0.11  & 8 & 6.6 $\pm$ 0.2 &  \\
\hline
O16 &  & 4.4 & 3.23 $\pm$ 0.05 & 13 & -70.8 $\pm$ 0.4 & {\multirow{3}{*}{199$\pm$6}} \\
&  & 4.9 & 2.96 $\pm$ 0.04 & 13 & -81.6  $\pm$  0.4 & \\
&  & 7.6 & 1.94 $\pm$ 0.05 & 13 & -105.3 $\pm$ 0.7 &  \\
\hline
O17 &  & 4.4 & 2.52 $\pm$ 0.03 & 10 & -27.2 $\pm$ 0.4 &  {\multirow{2}{*}{149$\pm$11}} \\
&  & 4.9 & 2.28 $\pm$ 0.03 & 10 & -34.7 $\pm$ 0.4 & \\
\enddata
\end{deluxetable}

\subsection{Comparison of Observations to HII Region Shell Models}
In this section, we compare our observations with mathematically simple expressions which describe the dynamics of an HII region interaction with the surrounding ISM. The first is the model presented in \citet{whi2009}. The \citet{whi2009} model contains a simple parameterization of a stellar bubble, as described by the theory of \citet{wea1977}. In that model, the HII region consists of an inner, low density cavity comprised of shocked stellar wind, and a contact discontinuity (assumed spherical) separating the shocked stellar wind from interstellar medium material. This interstellar medium material is shocked and photoionized interstellar gas which has passed through an outer shock. The last part of the bubble structure is the outer shock itself.\footnote{This model is described in more detail in Section 5.1 of \citet{whi2009}, and illustrated in Figure 6 of that paper.}
 The parameters of the model are R$_0$, the outer radius of the shell; R$_1$, the inner radius of the shell; $\emph{n$_{e}$}$, the plasma density within the shell (n$_{e}$=0 is assumed for r $<$ R$_1$); and $\vec{B_0}$, the interstellar magnetic field outside the shell. A distinction is made between the pristine magnetic field $\vec{B_0}$ upstream of the outer shock, and the downstream magnetic field inside the plasma shell, which has been modified by passage through the shock.

\citet{whi2009} obtain the following formula for the RM through such a shell.

\begin{equation}
RM(\xi)=\frac{Cn_{e}L(\xi)}{2}\left[B_{ZI}+B_{ZE}\right]
\label{rmxi}
\end{equation}
 where $n_e$ is the plasma density (electron density) in the shell, $L(\xi)$ is the length of the chord through the shell, and B$_{ZI}$ and B$_{ZE}$ are the downstream line of sight components of the magnetic field at the points where the line of sight enters (ingress) and leaves (egress) the shell respectively, given in Equations (7) - (9) of \citet{whi2009}. The variable $\xi$ is the transverse, linear distance between the line of sight and a line of sight passing through the center of the shell (i.e., $\xi$=0 is a line of sight through the center of the shell and $\xi$=R$_0$ is a line of sight which is tangent to the outer edge of the shell.).
The constant C is the collection of fundamental physical constants in curved brackets in Equation (\ref{RM1}). The constant C has the value 2.631$\times$10$^{-17}$ in cgs units, or 0.81 if ``interstellar units'' of cm$^{-3}$, $\mu$Gauss, and parsecs are chosen for n$_{e}$, $\vec{B_0}$, and L, respectively. L($\xi$) is given by
\begin{equation}
L(\xi)=2R_{0}\sqrt{(1-(\xi/R_0)^2)},  \mbox{ if } \xi \geq R_1
\end{equation}
\begin{displaymath}
L(\xi)=2R_{0}[\sqrt{(1-(\xi/R_0)^2)}-(R_1/R_0)\sqrt{(1-(\xi/R_1)^2)}],  \mbox{ if } \xi \leq R_1
\end{displaymath}

Exterior to the shell, we assume the magnetic field of the interstellar medium is uniform, but it will be modified in the shell. The theory of magnetohydrodynamic shock waves (e.g. \citet{gur2005}) shows that the magnetic field component in the shock plane is amplified by a factor X, and the component normal to the shock front is unchanged. The factor X, for the case of a strong shock, is equivalent to the density compression ratio. We redefine the B$_{ZI}$ and B$_{ZE}$ components in terms of B$_{0Z}$, the upstream line of sight component of the magnetic field. Employing these assumptions and definitions in Equation (\ref{rmxi}), we have
\begin{equation}
RM(\xi)=Cn_{e}L(\xi)B_{0Z} \left[ 1 + (X-1) \left( \frac{\xi}{R_{0}} \right)^2 \right]
\label{rmxi1}
\end{equation}
It should be pointed out that our shell model, expressed in Equations (6)-(8), assumes that the post-shock field strength at the ingress or egress point applies everywhere along the half-chord connecting the ingress or egress point to the midpoint of the chord (see Figure 6 of \cite{whi2009} for an illustration).  No attempt is made here to confront the physically complex question of the shell magnetic field as a function of position throughout the shell.  Our approximation is presumably accurate for a thin shell, in which the chord extends only a short distance from the shock front before entering the bubble interior (again, see Figure 6 of \cite{whi2009}).  However, in the case of a thick shell, this approximation must break down, and Equation (8) must be inaccurate.  Other than recognizing this fact, further investigation is beyond the scope of this paper.  This recognition should motivate further theoretical work to obtain analytic expressions which incorporate the results of MHD calculations such as \cite{fer1991} and \cite{sti2009}.
For our model of Equation (8), we adopt the shell parameters from \citet{cel1985}, where R$_0$= 16.9 parsecs, R$_1$= 6.2 parsecs, and $\emph{n$_{e}$}$= 10.8 - 15.5 cm$^{-3}$. These numbers refer to Celnik's Model 1, which is the single shell model. For the calculations described below we utilize a density equal to the mean of Celnik's values, $\emph{n$_{e}$}$=13.1 cm$^{-3}$. The variable B$_{0Z}$, the $z$ component of the upstream ISM magnetic field, is
\begin{displaymath}
 B_{0Z} = B_{0} \cos{\theta},
\end{displaymath}
where $B_{0}$ is the magnitude of the general interstellar field. In the analysis of this paper, we assume $B_{0}$ to be a known constant, and $\theta$ to be a variable with a wide range of possible values at a given point in the Galaxy.  The justification for this choice is the rather well established value for the magnitude of the magnetic field in the low density phases of the ISM \cite[e.g][]{fer2011,cru2010}. We choose $B_{0}$=4 $\mu$G in the calculations below. The angle $\theta$ may have a well-defined expectation value for the location of the Rosette Nebula in the Galaxy, but the actual value at a specific location and time presumably departs significantly from this expectation value due to turbulent fluctuations in the ISM.  A meaningful analogy would be the interplanetary magnetic field at 1 AU.  Although the average direction conforms to the Parker spiral, a measurement at a given time shows the field pointing in a wide range of directions.  

It should be recognized that in reality, both $B_{0}$ and $\theta$ are random variables with mean values and probability density functions.  As such, the true unknown variable is $B_{0z}$ which is formed from them.  Again, observations of the solar wind prove instructive.  Examination of several days of interplanetary magnetic field measurements show that the angles defining the direction of the interplanetary field show random variations, but the magnitude of the field does as well.  The solar wind provides some support for our practice in the present case.  Although the magnitude of the field does change with time, the fractional changes are usually relatively small in comparison with the large variations in the orientation of the interplanetary field.  This statement is supported by the well known observational result that the variance of the magnitude of the interplanetary field is much less than the variance in the components \citep{bru2005}.

\begin{figure}[!ht]
\begin{center}
\includegraphics[scale=0.66]{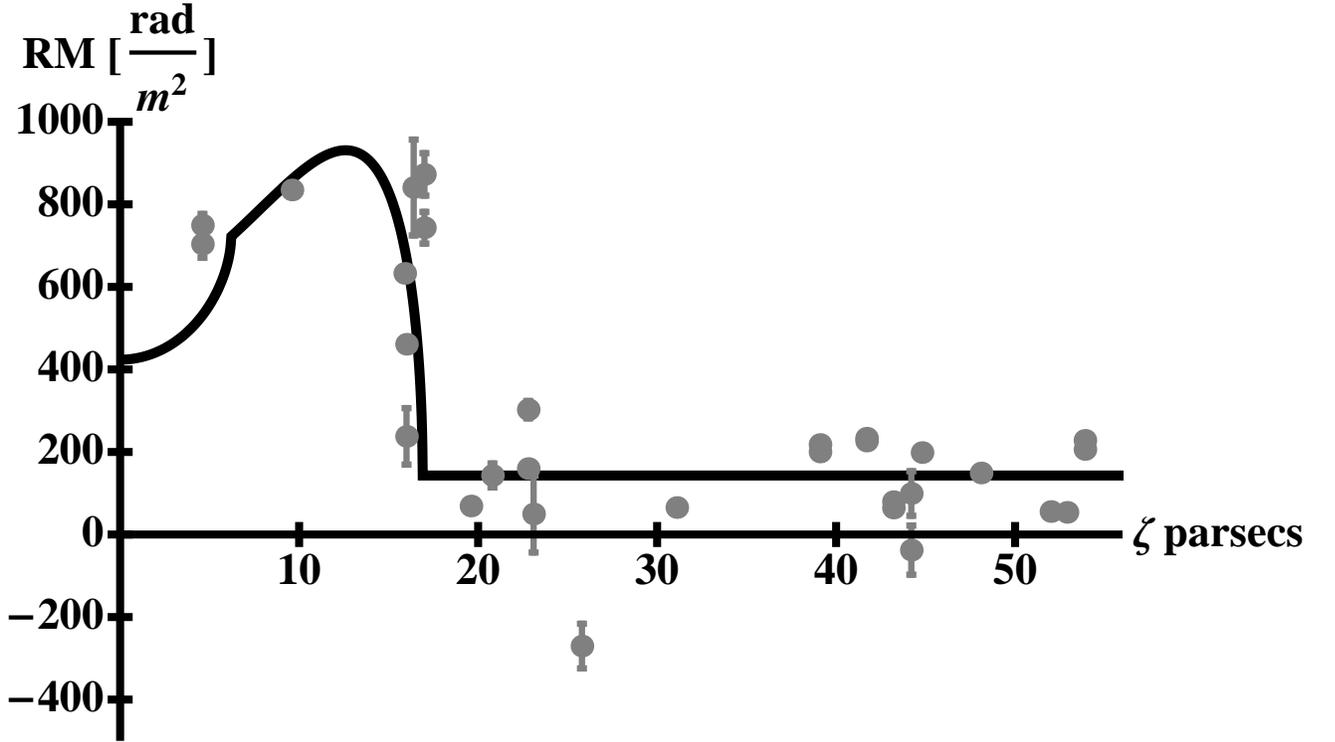}
\caption{Plot of RM versus distance from center of the Rosette Nebula. This plot differs from Figure \ref{figrmarc} in that the distance of the lines of sight from the nebular center have been converted from arcminutes to parsecs, and the model for Faraday rotation through a stellar bubble given by Equation (\ref{rmxi1}), has been overplotted. This model utilizes the following shell parameters: R$_1$=6.2 pc, R$_0$=16.9 pc, and n$_{e}$=13.1 cm$^{-3}$. Achieving this fit requires that the interstellar magnetic field at the location of the Rosette Nebula has a magnitude of 4 $\mu$G and is inclined at an angle $\theta$=72$^o$ with respect to our line of sight.}
\label{model1}
\end{center}
\end{figure}

\begin{figure}[!ht]
\begin{center}
\includegraphics[scale=0.66]{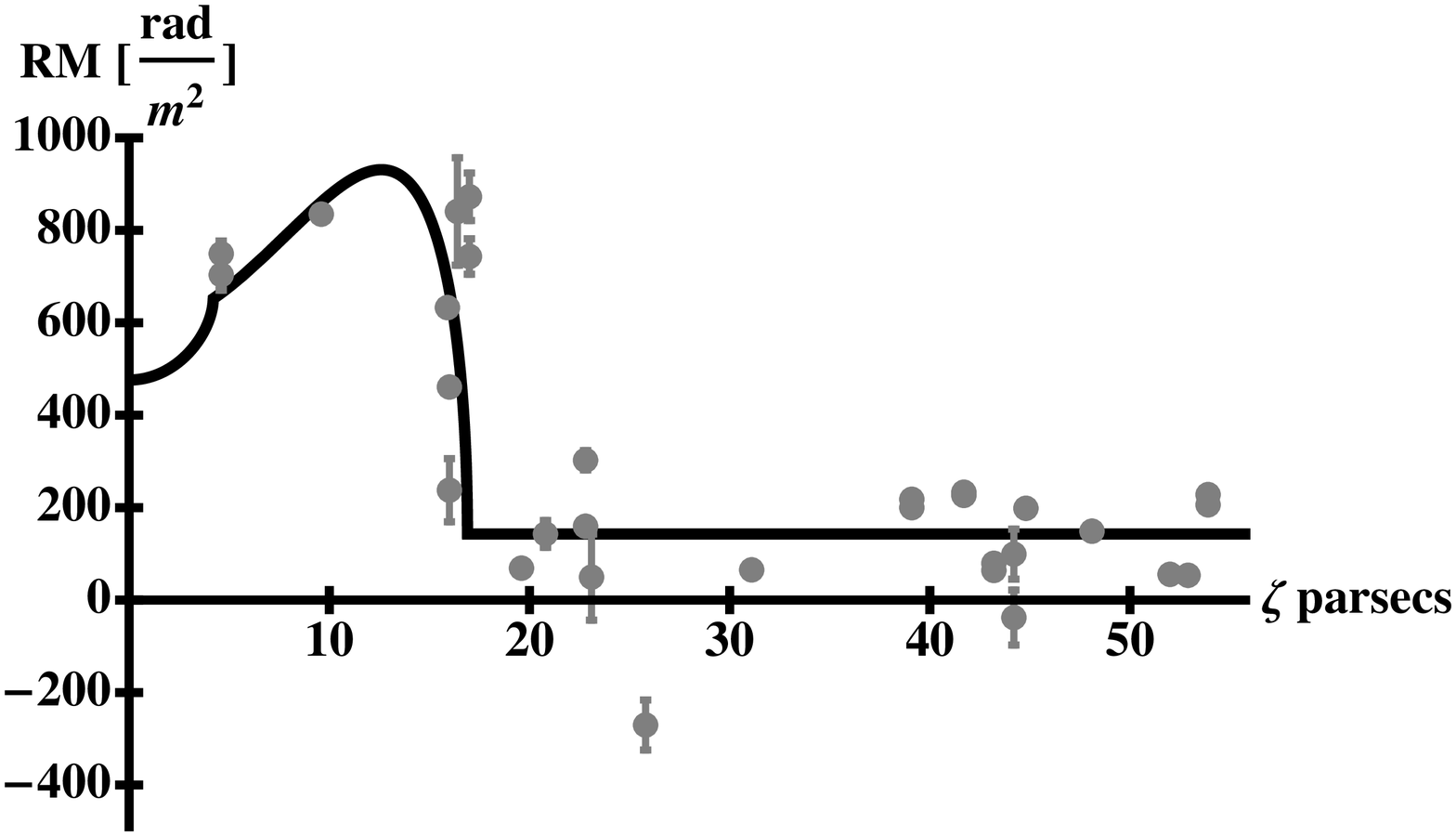}
\caption{This plot is the same as Figure \ref{model1} except the inner radius of the shell has been changed to optimize the fit to the data, R$_1$=4.2 pc. The value of $\theta$ is $\theta$ = 72$^o$.}
\label{fignewR}
\end{center}
\end{figure}

 In comparing the model of Equation (8) with our data, we overlaid curves generated by Equation (\ref{rmxi1}) on a plot of the RM vs the distance $\xi$ in parsecs from the center of the Rosette (Figure \ref{model1}), and effectively used the free parameter $\theta$ as a ``tuning knob'' for the  model.  By doing so, we obtained a value of $\theta$= 72$^{\circ}$ such that the model reproduces the magnitude of the measured RMs, and their dependence on the distance from the center of the Rosette Nebula. The degree of agreement between the model and the data in Figure \ref{model1} is actually quite good, particularly since we have adopted the shell model parameters R$_0$, R$_1$, and $\emph{n$_{e}$}$ directly from the data of \citet{cel1985}. We have not varied these parameters in an attempt to optimize the fit. Figure \ref{fignewR} presents the shell model with altered radii in order to obtain a better fit for the model Equation (\ref{rmxi1}) to the data.

To fit the magnitudes of the RMs viewed through the nebula, our model requires that the interstellar magnetic field at the location of the Rosette Nebula (before modification by the bubble associated with the Rosette) be rather highly inclined to the line of sight. Interestingly, our value of $\theta$ is roughly consistent with that expected for the mean Galactic field at the location of the Rosette Nebula. We use the galactic longitude of $206^{\circ}.5$ for the Rosette, and assume a Galactocentric distance of the Sun of 8.5 kpc, and a distance to the Rosette of 1.6 kpc.  In this case, the angle between the line of sight and an azimuthal magnetic field is $68^{\circ}$.  This is obviously completely consistent (to a doubtlessly fortuitous degree) with our model value of $\theta = 72^{\circ}$.  

Studies of the functional form of the Galactic magnetic field, while not conclusive at discriminating between an azimuthal field and one which follows the spiral arms, indicate that the field in the approximate neighborhood of the Sun has a pitch angle of $-8^{\circ}$ \citep{beck2001,fer2011}. Application of this pitch angle to an azimuthal field would then produce an expected angle of $60^{\circ}$ between the mean Galactic field at the location of the Rosette and the line of sight.  This value is also in acceptable agreement with our inferred value, in that it indicates a magnetic field that is oriented at a large angle with respect to the line of sight. 

\begin{figure}[!ht]
\begin{center}
\includegraphics[scale=0.57]{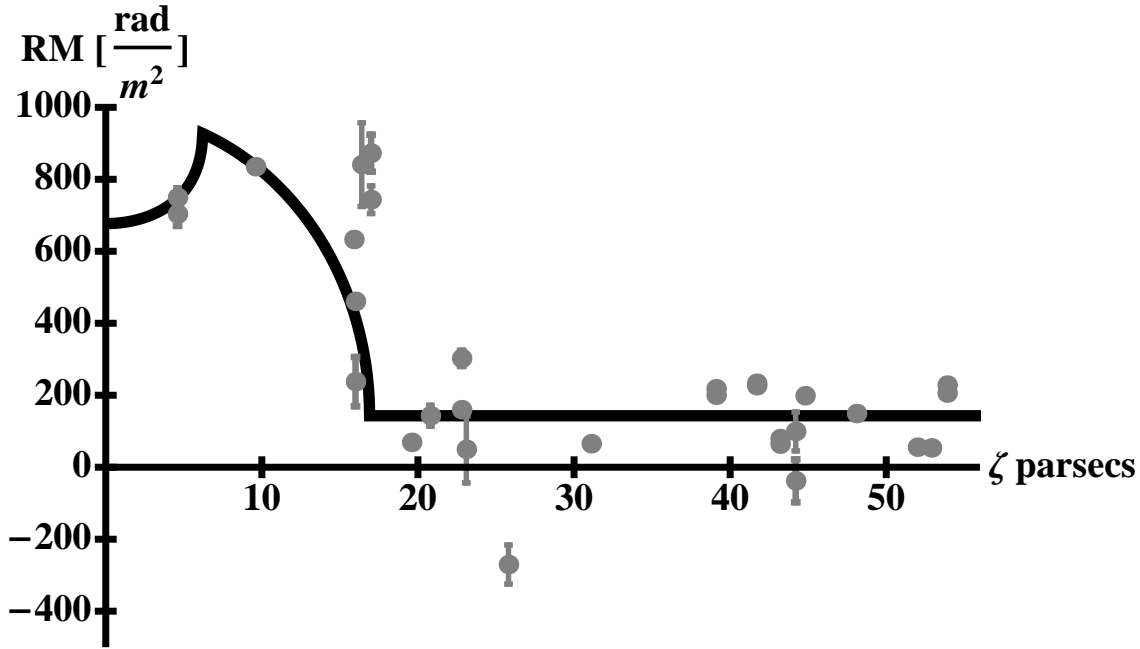}
\caption{This plot is the same as Figures \ref{model1} and \ref{fignewR} except the model that has been overplotted is given by Equation (\ref{rmxi2}) (interstellar magnetic field unmodified by HII region). This model curve requires that the interstellar magnetic field at the location of the Rosette Nebula is inclined at an angle $\theta$=54$^{\circ}$ with respect to our line of sight.}
\label{figrmmodels}
\end{center}
\end{figure}

We now consider a quite different HII region model which has been discussed in the context of Faraday rotation ``anomalies'', that of \citet{har2011} discussed in Section 1.3 above. \citet{har2011} concluded that the magnetic field was not amplified in the volume of the HII region. We have adjusted our Equation (\ref{rmxi}) to express the \citet{har2011} assumption of no $\vec{B}$ field amplification, giving the formula

\begin{equation}
RM(\xi)=Cn_{e}L(\xi)B_{0Z}
\label{rmxi2}
\end{equation}
where all parameters are defined following Equation (\ref{rmxi}). The difference between these two expressions is that Equation (\ref{rmxi2}) does not include amplification of the ``upstream'' interstellar magnetic field by the outer shock of the stellar bubble. As before, $\theta$ is the only free parameter and was varied to obtain a fit to the observed RM observations. By visual inspection, we obtained $\theta$=54$^{\circ}$ for  reasonable agreement between Equation (\ref{rmxi2}) and the data. A comparison of the model given by Equation (\ref{rmxi2}) with the data is shown in Figure \ref{figrmmodels}. Although it produces the magnitude and angular scale of the RM anomaly, it arguably does not do as well in reproducing the observed dependence of RM on distance from the center of the nebula. The smaller inclination of the interstellar magnetic field ($\theta$=54$^{\circ}$) is easily understood since in this latter model, there is no amplification of the perpendicular component of the interstellar magnetic field at the outer shock front (see Equation 9 of \citet{whi2009}). 
We suggest that the model of \citet{whi2009} provides a better fit to the observed dependence of RM on distance from the center of the nebula for the case of the Rosette Nebula. To distinguish between these two models will require more lines of sight which pass between the inner and outer radii of the bubble (6 and 17 parsecs in the case of the Rosette Nebula). For the shell models described by Equation (\ref{rmxi2}), in which the pre-existing interstellar magnetic field is unaltered by the presence of the HII region, the RM should have a maximum near the inner radius, as shown in Figure \ref{figrmmodels}. In the model of \citet{whi2009}, on the other hand, the magnetic field is amplified and ``refracted'' into the shock plane. This has the potential of producing ``RM limb brightening'', as might be present in the Rosette Nebula data, Figures \ref{model1} and \ref{fignewR}. This situation might be clarified by RM measurements of an additional 11 sources that were made with the VLA in February 2012, and are currently awaiting reduction and analysis.

The fit value for $\theta$ probably does not have much diagnostic ability, at least in the case of a single nebula. The values of $\theta$ for an azimuthal Galactic magnetic field (68$^{\circ}$), or a spiral field with a pitch angle of $-8^{\circ}$, $\theta = 60^{\circ}$ are more or less equally compatible with  our shell model or the unmodified field model, Equation (\ref{rmxi2}). However, all studies of the Galactic magnetic field show that the random component of the field is comparable to, if not larger than, the mean systematic component \citep[e.g.][]{rand1989,min1996, hav2006}. Thus the interstellar magnetic field at the location of the Rosette is doubtlessly comprised of a mean, large scale component which is inclined at a large angle to the line of sight, and a random, turbulent component which is isotropic. As a result, the ``local'' interstellar magnetic field at the Rosette Nebula could point in virtually any direction.

There are some final and obvious remarks which should be made regarding a comparison between the results of the present study of the Rosette Nebula and those of \citet{har2011} on 5 other HII regions. The model of \citet{whi2009} assumes that the plasma shell around the HII region is a bubble as described by the theory of \citet{wea1977}. The formation of a bubble on the scale of the Rosette Nebula requires stars with very large wind luminosities, which can only be furnished by very early main sequence stars or Wolf-Rayet stars. Such stars will only be found in very young stellar associations. At a later time, stellar wind luminosities will subside and the wind-blown bubble or superbubble will cease to exist. As was discussed in the Introduction, the Rosette Nebula is an excellent candidate for a stellar superbubble. The observations of \citet{men1962} showed that it has the annular shell structure expected of such a bubble, and it has been used as a paradigmatic wind-blown structure in theoretical studies [e.g. \citet{dor1986, dor1987}]. It therefore would be expected, $\emph{a priori}$, to show the plasma structure expected for a bubble.

The HII regions studied by \citet{har2011} could well be older clusters that are past the age when luminous stellar winds dominate their surroundings in the ISM. Resolution of this interesting question will require further observations of a sample of HII regions, with independent information on the ages of the star clusters and the wind luminosities of the constituent stars.

\subsection{Differences in Rotation Measure Between Closely-Spaced Lines 
of Sight}
The data in Table 3 show several cases in which RM is measurable for 
two components within the same source.  This raises the possibility of 
measuring RM differences between closely-spaced lines of 
sight.  \cite{Spangler07} uses the term {\em differential Faraday 
rotation} to describe such differences.  In the case of the present 
observations, with a synthesized beam width (FWHM) of 12\farcs8 and an 
assumed distance to the Rosette Nebula of 1600 pc, we can examine lines 
of sight separated by as little as 0.1 parsecs.

Differential Faraday rotation observations have been discussed by many 
authors \citep[e.g.][]{min1996,Haverkorn08}.  Measurements of the 
RM difference $\Delta RM$ on many pairs of lines of sight 
with a range of angular separations $\delta \theta$ can be used to 
construct the RM structure function $D_{RM}(\delta 
\theta)$.  The RM structure function yields 
characteristics of interstellar plasma turbulence 
\citep{min1996,Haverkorn08}.  \cite{Spangler07} also pointed out that a 
measurement of differential Faraday rotation could indicate the presence 
of an electrical current flowing between the lines of sight, and used a 
measurement of differential coronal Faraday rotation to deduce a 
model-dependent value for the magnitude of coronal currents.  The same 
ideas could be applied to measurements of interstellar Faraday rotation.

In this subsection, we briefly discuss the status of differential 
Faraday rotation in our sample of sources.  Nine of the sources in Table 
3 have RM values for two source components.  We restrict attention to 
those sources with $\chi$ measurements at three frequencies.  Such 
observations provide more secure and precise RM values.  These sources 
are I8, I14, O2, O4, O5, and O9.  Obviously, with such a restricted set 
of data we cannot construct a RM structure function, and 
our comments here will remain qualitative.

We first consider the four ``exterior'' sources O2, O4, O5, and O9.  The 
RM values for these sources are presumably determined by the general 
interstellar medium, with no contribution from the Rosette Nebula.  The 
$\Delta RM$ values for these sources range from $\sim 7 - 25$ rad m$^{-2}$, 
and in at least 2 cases (O5 and O9) seem consistent with zero, given the 
measurement errors.  The other two exterior sources (O2 and O4) have 
$\Delta RM$ values which appear slightly larger than expected for noise 
fluctuations about a zero expectation value.

The two ``interior'' sources I8 and I14 have $\Delta RM$ values in 
excess of 100 rad m$^{-2}$, and larger than expected from our error 
estimates.  This would seem to indicate enhanced differential Faraday 
rotation for lines of sight that pass through the interior of the 
Rosette Nebula, implying higher levels of plasma turbulence or 
electrical current systems flowing in the bubble associated with the 
Rosette.  However, two caveats should be noted.  First, as noted in Section 2, 
it is possible that component b of I14 is internally depolarized at the 
frequencies of observation, in which case neither the fit $RM$ for component b 
nor the measured $\Delta RM$ between the
two components is a diagnostic of the ISM.  Second, 
 the small number of sources we are considering (2 interior and 
4 exterior sources) precludes any firm conclusions about the statistics 
of differential Faraday rotation inside and outside of the Rosette 
Nebula.  

Given the data available in the present paper, a possible enhancement in 
$\Delta RM$ for the interior source I8 (and perhaps I14) relative to the 
exterior sources is speculative. The statistics of differential Faraday rotation for lines 
of sight passing through the Rosette Nebula, and the comparison with the 
statistics for lines of sight which do not pass through the nebula, need 
to be determined by measurements for a larger number of sources.  As 
mentioned in Section 4.1, multifrequency polarization measurements of an 
additional 11 sources with lines of sight through the nebula have been 
made and are awaiting reduction and analysis.  Those data should 
determine if an enhancement in differential Faraday rotation due to the 
Rosette Nebula exists, and if it does exist, establish its properties.

\section{Summary and Conclusions}
The conclusions of this paper are as follows.
\begin{enumerate}
\item{We observed 23 extragalactic radio sources whose lines of sight pass through or close to the Rosette Nebula and obtained Faraday rotation measurements for 21 of them. The interior sources, whose lines of sight pass through the Rosette, have an excess RM of 50-750 rad m$^{-2}$ with respect to a background due to this part of the galactic plane, which we determined to be +147 rad m$^{-2}$. We interpret this  50-750 rad m$^{-2}$ excess as the Faraday rotation measure of the plasma shell which comprises the Rosette Nebula.}
\item{We have compared our observations with a simplified analytic model for the plasma shell associated with a wind-driven, photoionized stellar bubble surrounding the NGC 2244 star cluster. This model was derived and presented in \citet{whi2009}. We find the measurements adhere well to the model if the angle between the line of sight and the Galactic magnetic field at the location of the Rosette Nebula is $\theta$=72$^{\circ}$ (see Figure \ref{model1}). This angle is compatible with that expected for the mean Galactic field at the location of the Rosette Nebula ($60^{\circ} - 68^{\circ}$). Our observations support an interpretation in which the Rosette Nebula is a wind-blown bubble as described by the theory of \citet{wea1977}.}
\item{We have also compared our observations with a simpler model in which the NGC 2244 star cluster photoionizes the surrounding gas without modifying the magnetic field, as proposed by \citet{har2011}. This model, unlike the stellar bubble model, does not naturally account for the observed, annular shell structure of the Rosette Nebula. This model can also reproduce the magnitude of the RMs measured through the Rosette Nebula, with a smaller angle between the line of sight and the interstellar field at the location of the Rosette ($\theta$=54$^{\circ}$). This model does not seem to account as well for the observed dependence of RM on the projected distance from the center of the nebula.}
\item{A determination of which of these models, if either, is better for the plasma structure of HII regions will require similar studies of more HII regions (with large numbers of lines of sight per HII region), spanning a range in age of the embedded star clusters.}
\item We have compared our $RM$ values with those of \cite{tay2009} for the 7 sources (with 10 source components) in common.  Good agreement between the two sets of measurements was found.  This comparison was principally undertaken as a check of the $RM$s resulting from our observations, but it also contributes to the literature on the accuracy of the large \cite{tay2009} $RM$ data set.  Our limited investigation supports the general accuracy of the \cite{tay2009} data, but does not contradict the finding of episodic inaccuracies or biases, as discussed in \cite{van2011}. 
\end{enumerate}

\acknowledgments{This research was supported at the University of Iowa by grant AST09-07911 and ATM09-56901 from the National Science Foundation. We are also grateful for advice and assistance from the staff of the National Radio Astronomy Observatory, Socorro, New Mexico. We thank the referee of this paper for a helpful and collegial review. We also thank Professors Marijke Haverkorn of Radboud University, Nijmegen, and Katia Ferri\`{e}re of the University of Toulouse for insightful, interesting, and helpful reviews of this paper. }

\end{document}